\def\arxiv{1}
\if\arxiv1
	\pdfoutput=1
\fi

\documentclass[10pt, journal, transmag]{IEEEtran}

\newcommand{\Papertitle}{Hybrid-State Free Precession in Nuclear Magnetic Resonance}
\newcommand{\linkcolor}{blue}

\usepackage{forloop}
\usepackage{xcolor}
\definecolor{UKLred} {RGB}{207, 25,  59}
\definecolor{UKLblue}{RGB}{ 47, 63, 157}
\definecolor{turquois}{rgb}{0,0.75,0.75}%

\usepackage[english]{babel}
\usepackage[sc]{mathpazo}
\usepackage[T1]{fontenc}
\usepackage{textcomp}

\usepackage{graphicx} 
\usepackage{stfloats}

\usepackage{amsmath}
\usepackage{amssymb}
\usepackage{amsthm}
\usepackage{amsfonts}
\usepackage{upgreek}

\usepackage{helvet}

\usepackage[helvet]{sfmath}

\usepackage{multirow}
\usepackage{dsfont}
\usepackage{widetext}

\usepackage[numbers, comma, super, sort&compress]{natbib}

\usepackage{colortbl}

\usepackage{hyperref}
\hypersetup{
	pdfpagelayout=TwoPageLeft,
	pdfstartview=Fit,
	bookmarksopen=true,
	bookmarksnumbered=true,
	pdftitle={\Papertitle},
	pdfauthor={Jakob Assländer},
	pdfsubject={Manuscript},
	breaklinks=true,
	colorlinks=true,
	linkcolor=\linkcolor,
	anchorcolor=black,
	citecolor=\linkcolor,
	filecolor=black,
	menucolor=red,
	urlcolor=\linkcolor,
	pdfencoding=auto
}

\usepackage{xr}
\usepackage[font={footnotesize}]{caption}
	
\usepackage{animate}
\usepackage{tikz}
\usepackage{pgfplots}
\pgfplotsset{compat=1.14,
	colormap={parula}{
		rgb255=(53,42,135)
		rgb255=(15,92,221)
		rgb255=(18,125,216)
		rgb255=(7,156,207)
		rgb255=(21,177,180)
		rgb255=(89,189,140)
		rgb255=(165,190,107)
		rgb255=(225,185,82)
		rgb255=(252,206,46)
		rgb255=(249,251,14)},
	colormap={mybluered}{
		rgb255(0cm)=(0,0,180)
		rgb255(1cm)=(0,180,180)
		rgb255(2cm)=(70,180,0)
		rgb255(3cm)=(180,180,0)
		rgb255(4cm)=(255,0,0)
		rgb255(5cm)=(128,0,0)}
	}

\usetikzlibrary{spy,backgrounds,arrows,decorations.pathmorphing,backgrounds,positioning,fit,matrix,calc}
\pgfdeclarelayer{background}
\pgfdeclarelayer{foreground}
\pgfsetlayers{background,main,foreground}

\RequirePackage{shellesc}
\usepgfplotslibrary{external}
\usepackage[shortcuts]{extdash}

\usepackage[symbol]{footmisc}

\include{Figures/blochsphere}

\begin{document}
\graphicspath{{Figures/}} 

\title{\Papertitle}

\author{\IEEEauthorblockN{\bf \normalsize Jakob Assl\"ander$^{1,2,\ast}$, Dmitry S. Novikov$^{1,2}$, Riccardo Lattanzi$^{1,2,3}$, Daniel K. Sodickson$^{1,2,3}$, Martijn A. Cloos$^{1,2}$}
	
	\IEEEauthorblockA{\small$^1$ Center for Biomedical Imaging, Dept. of Radiology, New York University School of Medicine, New York, NY, USA}
	
	\IEEEauthorblockA{\small$^2$ Center for Advanced Imaging Innovation and Research, New York University School of Medicine, New York, NY, USA}
	
	\IEEEauthorblockA{\small$^3$ The Sackler Institute of Graduate Biomedical Sciences, New York University School of Medicine, New York, NY, USA \vspace{-1.5cm}}
}
\maketitle

\noindent
\sloppy
\begin{abstract}	
	The dynamics of large spin-1/2 ensembles in the presence of a varying magnetic field are commonly described by the Bloch equation. Most magnetic field variations result in unintuitive spin dynamics, which are sensitive to small deviations in the driving field. Although simplistic field variations can produce robust dynamics, the captured information content is impoverished. 
	Here, we identify adiabaticity conditions that span a rich experiment design space with tractable dynamics.
	These adiabaticity conditions trap the spin dynamics in a one-dimensional subspace. Namely, the dynamics is captured by the absolute value of the magnetization, which is in a transient state, while its direction adiabatically follows the steady state. We define the \textit{hybrid state} as the co-existence of these two states and identify the polar angle as the effective driving force of the spin dynamics.
	As an example, we optimize this drive for robust and efficient quantification of spin relaxation times and utilize it for magnetic resonance imaging of the human brain. 
\end{abstract}

For many nuclei, the spin gives rise to a magnetic moment, whose dynamics can be used for quantum computing  \cite{Gershenfeld1997} and provides a window to study, e.g., the chemical structure of molecules, as done in nuclear magnetic resonance \cite{Rabi1938} (NMR) spectroscopy, or the composition of biological tissue, as used for clinical diagnosis in magnetic resonance imaging \cite{Lauterbur1973} (MRI). 
Modeling spin-lattice and spin-spin interactions as random magnetic field fluctuations \cite{Bloembergen1948} allows for capturing their macroscopic effect by the relaxation times $T_1$ and $T_2$, respectively. This facilitates the description of large spin-1/2 ensembles with the classical Bloch equation \cite{Bloch1946}, formally akin to the time-dependent Schrödinger Equation in a 4D-space:
\begin{equation}
\partial_t \begin{pmatrix} x \\ y \\ z \\ 1 \end{pmatrix} = 
\begin{pmatrix} 
-\frac{1}{T_2} & -\omega_z       & \omega_y & 0  \\ 
\omega_z       & -\frac{1}{T_2}   & -\omega_x & 0  \\
-\omega_y     & \omega_x         & -\frac{1}{T_1} & \frac{1}{T_1} \\
0 & 0 & 0 & 0
\end{pmatrix} 
\begin{pmatrix} x \\ y \\ z \\ 1 \end{pmatrix}.
\label{eq:Bloch}
\end{equation}
Here, $\partial_t$ denotes the partial derivative with respect to time, $x, y, z$ are the spatial components of the magnetization, and $1$ is the normalized $z$-magnetization at thermal equilibrium. The Rabi frequencies \cite{Rabi1938} $\omega_x$ and $\omega_y$ (induced by radio frequency (RF) pulses), together with the Larmor frequency $\omega_z$, are the external drive of the spin dynamics. 

While the Bloch equation is very general, it provides little intuition to help design robust and efficient experiments. 
This lack of intuition has biased experimental design towards elementary drives for which analytic solutions make the effect of spin relaxation and experimental imperfections evident. For example, the workhorses of clinical MRI weight the signal intensity either by $T_1$ or $T_2$ effects by exploiting the simplest spin dynamics, most notably exponential relaxation \cite{Hahn1950,Hennig1986,Mugler1990} and steady states \cite{Carr1958,Haase1986,Oppelt1986}. These basic drives span small subspaces like the steady-state ellipse \cite{Carr1958, Freeman1971, Hennig2002, Lapert2013},  which harbor impoverished spin dynamics compared to the richness found outside.
More recent approaches strive to break away from such traditional experimental design in search for an improved signal-to-noise efficiency \cite{Ma2013}. However, the non-intuitive nature of the Bloch equation has limited the exploration of this vast experiment design space to heuristic guesses \cite{Ma2013, Jiang2015, Cloos2016, Ma2016, Asslander2017, Jiang2017}. 

The rationale for this improved encoding efficiency is sketched in Fig.~\ref{fig:MRF_HSFP_SSFP}: Variations of the driving fields result in a transient state, which enables one to exploit the entire Bloch sphere in search for the optimal encoding of characteristic parameters such as spin relaxation times. The same plot also points out a risk associated with the transient state: Small magnetic field deviations can produce substantially differing spin trajectories, which can bias the estimation of characteristic parameters. 
This is particularly problematic in biological tissue, where inhomogeneous broadening is inevitable and difficult to model \cite{Asslander2017,Ganter2006}. 

Here, we formulated conditions under which the sensitivity to magnetic field deviations and inhomogeneous broadening is greatly mitigated and reveal a large subspace of drives in which the Bloch equation is tractable. Our analysis shows that, under these conditions, the direction of the magnetization adiabatically follows the one of steady states, while the absolute value of the magnetization can be in a transient state. 
In this \textit{hybrid state}, the spin dynamics live, therefore, in a one-dimensional subspace and can be described by a 2x2 Hamiltonian:
\begin{equation}
\partial_t \begin{pmatrix} r \\ 1 \end{pmatrix} = \begin{pmatrix} -\frac{\cos^2 \vartheta}{T_1} -\frac{\sin^2 \vartheta}{T_2} & \frac{\cos \vartheta}{T_1} \\ 0 & 0 \end{pmatrix} \begin{pmatrix} r \\ 1 \end{pmatrix},
\label{eq:Bloch_2D}
\end{equation}
where $r$ is the magnetization along the radial direction, i.e. its magnitude (cf. Section \ref{sec:Bloch_Spherical} for the derivation). 
This notation identifies the polar angle $\vartheta(t)$, which is the angle between the $z$-axis and the magnetization, as the relevant degree of freedom, which describes the joint effect of the drives $\omega_{x}(t)$, $\omega_{y}(t)$, and $\omega_{z}(t)$ on the spin dynamics. 
As an example, we show that this hybrid-state equation and its solution provide intuition for the encoding processes of spin relaxation times and are an excellent basis for numerical optimizations of a $T_1$ and $T_2$ mapping experiment that combines the robustness of the steady state with the encoding efficiency of the transient state. 

\section{Hybrid State Boundary Conditions}
\begin{figure*}[tbp]
	\centering
	\begin{tikzpicture}[scale = 1]
	\def\viewx{45}
	\def\viewy{20}
	
	\def\coord{
		(-6, -3, -1) [0.00]
		(-6, -3, 0) [1.00]
		(-6, -3, 1) [0.53]
		(-6, 0, -1) [0.57]
		(-6, 0, 0) [0.41]
		(-6, 0, 1) [0.01]
		(-6, 3, -1) [0.70]
		(-6, 3, 0) [0.51]
		(-6, 3, 1) [1.00]
		(-6, 6, -1) [0.06]
		(-6, 6, 0) [0.36]
		(-6, 6, 1) [0.23]
		(-3, -3, -1) [0.90]
		(-3, -3, 0) [0.81]
		(-3, -3, 1) [0.39]
		(-3, 0, -1) [0.05]
		(-3, 0, 0) [0.38]
		(-3, 0, 1) [0.77]
		(-3, 3, -1) [0.17]
		(-3, 3, 0) [0.91]
		(-3, 3, 1) [0.32]
		(-3, 6, -1) [0.33]
		(-3, 6, 0) [0.20]
		(-3, 6, 1) [0.77]
		(0, -3, -1) [0.07]
		(0, -3, 0) [0.95]
		(0, -3, 1) [0.16]
		(0, 0, -1) [0.29]
		(0, 0, 0) [0.69]
		(0, 0, 1) [0.14]
		(0, 3, -1) [0.51]
		(0, 3, 0) [0.72]
		(0, 3, 1) [0.93]
		(0, 6, -1) [0.73]
		(0, 6, 0) [0.75]
		(0, 6, 1) [0.41]
		(3, -3, -1) [0.24]
		(3, -3, 0) [0.52]
		(3, -3, 1) [0.22]
		(3, 0, -1) [0.84]
		(3, 0, 0) [0.66]
		(3, 0, 1) [0.82]
		(3, 3, -1) [0.15]
		(3, 3, 0) [0.47]
		(3, 3, 1) [0.31]
		(3, 6, -1) [0.69]
		(3, 6, 0) [0.00]
		(3, 6, 1) [0.77]
	}
	
	\pgfdeclareplotmark{randspin}{
		\def\myphi{\pgfplotspointmeta*180}
		\def\xshift{-7.5}
		\def\zshift{-1.75}
		\draw[->, thick, color=black] ({\xshift-0.4*cos(\myphi)},{-0.4*sin(\myphi)},{\zshift-0.3}) -- ({\xshift+0.4*cos(\myphi)},{0.4*sin(\myphi)},{\zshift+0.4});
	      \shade[draw=none,shading=ball] (\xshift,0,\zshift) circle (\pgfplotmarksize);
	}

	\pgfdeclareplotmark{sortedspin}{
	\def\myphi{0}
	\def\xshift{-7.5}
	\def\zshift{-1.75}
	\draw[->, thick, color=black] ({\xshift-0.4*cos(\myphi)},{-0.4*sin(\myphi)},{\zshift-0.3}) -- ({\xshift+0.4*cos(\myphi)},{0.4*sin(\myphi)},{\zshift+0.4});
	\shade[draw=none,shading=ball] (\xshift,0,\zshift) circle (\pgfplotmarksize);
}

	\blochsphere{name=transient_B0, title = {transient state}}{
		\addplot3 [mesh,  mesh/cols = 1, point meta=explicit, z buffer=sort]
		table[x=Mx, y=My, z=Mz, meta=w]{Figures/Bloch_dynamcis_rand_B0.txt};
	}

	\begin{axis}[
		width = 4cm,
		at=(transient_B0.center),
		anchor=south east,
		xshift = -0.25cm,
		xmin = -6, xmax = 6,
		ymin = -6, ymax = 6,
		zmin = -1, zmax = 1.5,
		xtick=\empty, ytick=\empty, ztick=\empty,
		colormap/bluered,
		]
		
		\begin{pgfonlayer}{background}
		\fill[gray!90] (-6,-6,-1) -- (-6,6,-1) -- (6,6,-1) -- (6,-6,-1) -- cycle; 
		\fill[gray!90] (-6,-6,-1) -- (-6,6,-1) -- (-6,6,1.5) -- (-6,-6,1.5) -- cycle; 
		\fill[gray!90] (-6,6,-1) -- (6,6,-1) -- (6,6,1.5) -- (-6,6,1.5) -- cycle; 
		\end{pgfonlayer}{background}
		
		\begin{pgfonlayer}{foreground}
		\draw[black] (-6,-6,1.5) -- (-6,6,1.5) -- (6,6,1.5) -- (6,-6,1.5) -- cycle; 
		\draw[black] (-6,-6,-1) -- (6,-6,-1) -- (6,-6,1.5) -- (-6,-6,1.5) -- cycle;
		\end{pgfonlayer}{foreground}
		
		\addplot3+[
			z buffer=sort,
			scatter/use mapped color={ball color=mapped color},
			scatter,
			scatter src=explicit,
			only marks,
			mark=randspin,
			mark size=2pt,
		] coordinates{\coord};
	\end{axis}
	
	\blochsphere{name=HSFP_B0, at=(transient_B0.right of north east), anchor=left of north west, xshift = 1.5cm, title = {hybrid state}}{
		\addplot3 [mesh,  mesh/cols = 1, point meta=explicit, z buffer=sort]
		table[x=Mx, y=My, z=Mz, meta=w]{Figures/Bloch_dynamcis_HSFP_B0.txt};
	}

	\begin{axis}[
		width = 4cm,
		at=(HSFP_B0.center),
		anchor=south east,
		xshift = -0.25cm,
		xmin = -6, xmax = 6,
		ymin = -6, ymax = 6,
		zmin = -1, zmax = 1.5,
		xtick=\empty, ytick=\empty, ztick=\empty,
		colormap/bluered,
		]
		
		\begin{pgfonlayer}{background}
		\fill[gray!90] (-6,-6,-1) -- (-6,6,-1) -- (6,6,-1) -- (6,-6,-1) -- cycle; 
		\fill[gray!90] (-6,-6,-1) -- (-6,6,-1) -- (-6,6,1.5) -- (-6,-6,1.5) -- cycle; 
		\fill[gray!90] (-6,6,-1) -- (6,6,-1) -- (6,6,1.5) -- (-6,6,1.5) -- cycle; 
		\end{pgfonlayer}{background}
		
		\begin{pgfonlayer}{foreground}
		\draw[black] (-6,-6,1.5) -- (-6,6,1.5) -- (6,6,1.5) -- (6,-6,1.5) -- cycle; 
		\draw[black] (-6,-6,-1) -- (6,-6,-1) -- (6,-6,1.5) -- (-6,-6,1.5) -- cycle;
		\end{pgfonlayer}{foreground}
		
		\addplot3+[
			z buffer=sort,
			scatter/use mapped color={ball color=mapped color},
			scatter,
			scatter src=explicit,
			only marks,
			mark=sortedspin,
			mark size=2pt,
		] coordinates{\coord};
	\end{axis}
	
	\blochsphere{name=SSFP_B0, at=(HSFP_B0.right of north east), anchor=left of north west, xshift = 1.5cm, colorbar, colorbar style={height = 3cm, width = 0.3cm, ylabel = $\phi/\pi$, yshift = -1.5cm}, title = {steady state}}{
		\addplot3 [mesh,  mesh/cols = 1, point meta=explicit, z buffer=sort]
		table[x=Mx, y=My, z=Mz, meta=w]{Figures/Bloch_dynamcis_SSFP_B0.txt};
	}

	\begin{axis}[
		width = 4cm,
		at=(SSFP_B0.center),
		anchor=south east,
		xshift = -0.25cm,
		xmin = -6, xmax = 6,
		ymin = -6, ymax = 6,
		zmin = -1, zmax = 1.5,
		xtick=\empty, ytick=\empty, ztick=\empty,
		colormap/bluered,
		]
		
		\begin{pgfonlayer}{background}
		\fill[gray!90] (-6,-6,-1) -- (-6,6,-1) -- (6,6,-1) -- (6,-6,-1) -- cycle; 
		\fill[gray!90] (-6,-6,-1) -- (-6,6,-1) -- (-6,6,1.5) -- (-6,-6,1.5) -- cycle; 
		\fill[gray!90] (-6,6,-1) -- (6,6,-1) -- (6,6,1.5) -- (-6,6,1.5) -- cycle; 
		\end{pgfonlayer}{background}
		
		\begin{pgfonlayer}{foreground}
		\draw[black] (-6,-6,1.5) -- (-6,6,1.5) -- (6,6,1.5) -- (6,-6,1.5) -- cycle; 
		\draw[black] (-6,-6,-1) -- (6,-6,-1) -- (6,-6,1.5) -- (-6,-6,1.5) -- cycle;
		\end{pgfonlayer}{foreground}
		
		\addplot3+[
			z buffer=sort,
			scatter/use mapped color={ball color=mapped color},
			scatter,
			scatter src=explicit,
			only marks,
			mark=sortedspin,
			mark size=2pt,
		] coordinates{\coord};
	\end{axis}
	
	\blochsphere{name=transient_B1, at=(transient_B0.below south), anchor=north, yshift = 0.75cm}{
		\addplot3 [mesh,  mesh/cols = 1, point meta=explicit, z buffer=sort]
		table[x=Mx, y=My, z=Mz, meta=B1]{Figures/Bloch_dynamcis_rand_B1.txt};
	}
	
	\blochsphere{name=HSFP_B1, at=(transient_B1.right of north east), anchor=left of north west, xshift = 1.5cm}{
		\addplot3 [mesh,  mesh/cols = 1, point meta=explicit, z buffer=sort]
		table[x=Mx, y=My, z=Mz, meta=B1]{Figures/Bloch_dynamcis_HSFP_B1.txt};
	}
	
	\blochsphere{name=SSFP_B1, at=(HSFP_B1.right of north east), anchor=left of north west, xshift = 1.5cm, colorbar, colorbar style={height = 3cm, width = 0.3cm, ylabel = $B_1 / B_1^{\text{nominal}}$, yshift = -1.5cm}}{
		\addplot3 [mesh,  mesh/cols = 1, point meta=explicit, z buffer=sort]
		table[x=Mx, y=My, z=Mz, meta=B1]{Figures/Bloch_dynamcis_SSFP_B1.txt};
	}
	\end{tikzpicture}
	\caption{In the fully-transient state (here visualized on the left for the example of a random RF-pattern), the spin trajectories on the Bloch sphere are, in general, very sensitive to magnetic field inhomogeneities. Deviations of the Larmor frequency are depicted in the top row, where $\phi$ denotes the phase accumulated over one \textit{repetition time} $T_R$, and we define $\phi = \pi$ as the on-resonance condition. The bottom row sketches spin trajectories for deviations in the RF field $B_1$, which alter the Rabi-frequencies. The signal of an NMR sample or a volume element in MRI (visualized by the cube) is generated by spins at different Larmor frequencies, which additionally introduces a strong sensitivity of the signal to the particular distribution of Larmor frequencies \cite{Ganter2004}. The hybrid state, shown at center, is explicitly designed to mitigate these sensitivities, while still allowing the magnetization to visit the entire Bloch sphere. 
	Fully adiabatic transitions between steady states, shown at right, have the same robustness to magnetic field deviations, however, they trap the magnetization on the steady-state ellipse \cite{Carr1958,Freeman1971,Hennig2002,Lapert2013}, which diminishes the capabilities to encode tissue properties such as relaxation times. The steady-state ellipse is described by setting the left hand side of Eq.~\eqref{eq:Bloch_2D} to zero. 
	}
	\label{fig:MRF_HSFP_SSFP}
\end{figure*}
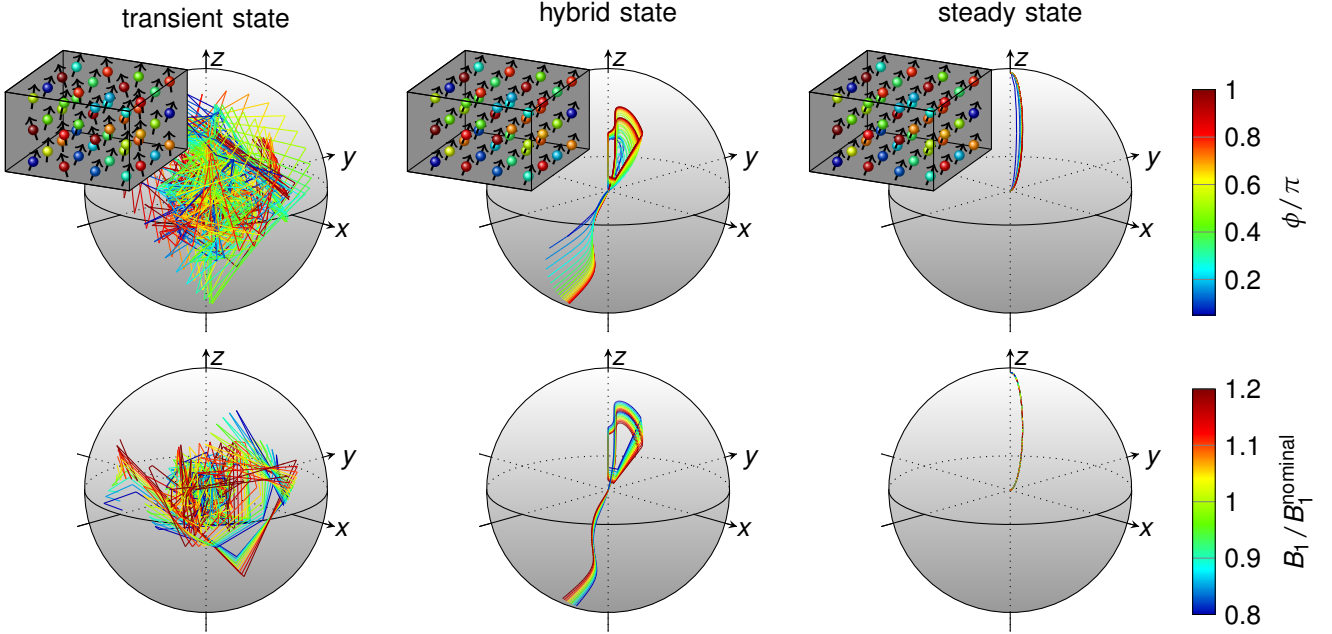

As the magnetization described by Eq.~\eqref{eq:Bloch} is real-valued, we can conclude that the eigenvalues of the Hamiltonian must either be real-valued or occur in complex conjugate pairs. One eigenvalue is zero and describes the steady-state magnetization. Therefore, another eigenvalue must be real-valued. As such, it describes an exponential decay of the corresponding transient-state component, while the remaining complex eigenvalues describe oscillatory decays. 
Ganter pointed out that the complex phase makes the latter components very sensitive to deviations in the magnetic field and in particular to inhomogeneous broadening \cite{Ganter2004}. Fig.~\ref{fig:MRF_HSFP_SSFP} provides some intuition for this sensitivity: As the complex phase accumulates during the experiment, the spin trajectory becomes very sensitive to deviations in the magnetic fields. Considering that the measured signal is invariably given by the integral over some distribution of Larmor frequencies, which is difficult to model in biological tissue \cite{Ganter2006}, contributions of the complex eigenvalues will lead to a bias in the estimated relaxation parameters \cite{Asslander2017}. 

Conversely, if we design our MR experiment such that the cumbersome complex eigenstates are not populated, we achieve robustness to magnetic field deviations and inhomogeneous broadening. If we simultaneously populate the real-valued transient eigenstate, we liberate the magnetization from the steady-state ellipse and gain access to the entire Bloch sphere (Fig.~\ref{fig:MRF_HSFP_SSFP}). 

In general, variations of the driving fields rotate the eigenvectors and populate all transient eigenstates. A Taylor expansion of this eigenbasis rotation (cf. Section \ref{sec:Adiabaticity}) reveals that this population is dominated by the gaps between the eigenvalue and the rest of the Hamiltonian's spectrum, similar to the quantum mechanical adiabatic theorem \cite{Born1928}. 
The real-valued eigenvalue is close to the steady-state eigenvalue, resulting in a very restrictive boundary condition. On the contrary, the complex eigenvalues are well separated from the rest of the spectrum due their complex phase, resulting in a less restrictive boundary condition. 

For pulsed experiments \cite{Hahn1950}, which dominate modern MR, we find the condition
\begin{equation}
\max \{ |\Delta \alpha|, |\Delta \phi| \} \ll \sin^2 \frac{\alpha}{2} + \sin^2 \frac{\phi}{2} - \frac{5}{2} \left( 1 - E_2 \right)
\label{eq:adiabaticity_HSFP}
\end{equation}
under which the complex eigenstates are not populated, and 
\begin{equation}
\max \{ |\Delta \alpha|, |\Delta \phi| \} \ll (1 - E_1)^2
\label{eq:adiabaticity_SSFP}
\end{equation}
under which the real-valued eigenstate is not populated. 
Here, the driving fields are parameterized by the flip angle $\alpha$ and the accumulated phase $\phi = \omega_z T_R$, where the \textit{repetition time} $T_R$ denotes the time between consecutive RF pulses, and $\Delta \alpha$ and $\Delta \phi$ denote the change of these parameters in consecutive repetitions. Relaxation is described by $E_{1,2} = \exp (- T_R/T_{1,2})$. 

Experiments in which Eq.~\eqref{eq:adiabaticity_HSFP} holds, but Eq.~\eqref{eq:adiabaticity_SSFP} does not, result in non-trivial, yet tractable spin dynamics that are rich in information content. 
Since the latter adiabaticity condition is substantially more restrictive, the \textit{hybrid state} theory governs a vast experiment design space. 
In order to provide some intuition, we can assume $T_R = 4.5$~ms, and relaxation times of human brain white matter ($T_1 = 781$~ms and $T_2 = 65$~ms) \cite{Jiang2015}. In such a case, $\max \{ |\Delta \alpha|, |\Delta \phi| \} \ll 1$ suffices to avoid a population of the complex eigenstates when, e.g., assuming $\phi = \pi$. In contrast, $\max \{ |\Delta \alpha|, |\Delta \phi| \} \ll 10^{-5}$ would be required to avoid a population of the real-valued transient eigenstate. 

\section{Adiabaticity and the Solution of the Bloch Equation}
Hargreaves et al. showed that the eigenvector corresponding to the complex eigenvalue is approximately perpendicular to the steady-state magnetization \cite{Hargreaves2001}, while the real-valued eigenvalue describes the transient-state component parallel to the steady-state magnetization. By enforcing Eq.~\eqref{eq:adiabaticity_HSFP}, we, thus, effectively force the direction of the magnetization to adiabatically follow that of the steady states. 
If we then simultaneously pick our driving fields to violate Eq.~\eqref{eq:adiabaticity_SSFP}, the magnitude of the magnetization is in a transient state, and a hybrid of two co-existing states emerges, which we dub \textit{hybrid state}.

The adiabaticity of the magnetization's direction effectively decouples the components of the Bloch equation, which allows us to formulate an analytic solution. For this purpose, we transform the Bloch equation into spherical coordinates and provide the solutions for the polar angle $\vartheta$, the phase $\varphi$, and the radius $r$, which we here define as the magnitude combined with a sign (cf. Section~\ref{sec:Bloch_Spherical} for the derivation). 
Except in the vicinity of the stop bands, which are defined by $|\sin \phi| \ll 1$ (cf. supporting Fig.~\ref{fig:Spin_Dynamics}), the polar angle can be approximated by
\begin{equation}
\sin^2 \vartheta = \frac{\sin^2 \frac{\alpha}{2}}{\sin^2 \frac{\phi}{2}  \cdot \cos^2 \frac{\alpha}{2} + \sin^2 \frac{\alpha}{2}}.
\label{eq:vartheta}
\end{equation}
This equation reduces to $\vartheta = \alpha/2$ for $\phi = \pi$, which we define as the on-resonance condition. In practice, $\phi = \pi$ is assigned to the on-resonant spin isochromat by the common phase increment of $\pi$ in consecutive RF pulses. 
The phase of the magnetization is approximated by
\begin{equation}
\varphi = \tan^{-1} \left( \frac{\cos \phi - E_2}{\sin \phi} \right) - \mathcal{H} \{\sin \phi \} \cdot \pi + \phi_{T_E},
\label{eq:varphi}
\end{equation}
where the Heaviside function $\mathcal{H}$ disambiguates the four-quadrants and $\phi_{T_E}$ describes the phase of the magnetization accumulated between the RF pulse and the time the signal is observed, i.e.,  the echo time $T_E$. 

The radial component $r$ captures the entire spin dynamics, which is described by a single first order differential equation (Eq.~\eqref{eq:Bloch_2D}). This equation is solved by 
\begin{equation}
	r(t) = a(t) \cdot \left(r(0) + \frac{1}{T_1} \int_{0}^{t} \frac{\cos \vartheta(\tau)}{a(\tau)} d\tau \right)
	\label{eq:r}
\end{equation}
with
\begin{equation*}
	a(\tau) = \exp \left( - \int_{0}^{\tau} \frac{\sin^2\vartheta(\xi)}{T_2} + \frac{\cos^2\vartheta(\xi)}{T_1} d\xi \right).
\end{equation*}
Here, $t$ denotes time and $r(0)$ the initial magnetization. 
Alternatively, we can define the initial magnetization as a function of the final magnetization, i.e. $r(0) = \beta \cdot r(T_C)$, where $T_C$ denotes the duration of a single cycle of the experiment. With this boundary condition, the radial Bloch equation is solved by
Eq.~\eqref{eq:r} with
\begin{equation*}
r(0) = \frac{\beta}{T_1} \frac{a(T_C)}{1 - \beta a(T_C)}  \int_{0}^{T_C} \frac{\cos \vartheta(\tau)}{a(\tau)} d\tau .
\end{equation*}
When we set $\beta = 1$, a periodic boundary condition is obtained, which requires the magnetization at the beginning and the end of each cycle to be equal. Similarly, $\beta = -1$ leads to an anti-periodic boundary condition, which implies an inversion of the magnetization between cycles. Such boundary conditions enable the concatenation of multiple cycles without delays, thus, allowing for efficient signal averaging and a flexible implementation, e.g., of time-consuming 3D imaging experiments. 

Intuitively, Eq. \eqref{eq:r} describes a predominant $T_1$ encoding at small $\vartheta$-values (close to the $z$-axis), and a predominant $T_2$ encoding as $\vartheta$ approaches $\pi/2$, which corresponds to the $x$-$y$-plane. When $\vartheta$ is constant, Eq.~\eqref{eq:r} reduces to the exponential transition into steady state described by Schmitt et al. \cite{Schmitt2004} (cf. supporting material). 

Supporting Fig.~\ref{fig:Spin_Dynamics} validates the hybrid-state model by comparing Eqs.~\eqref{eq:vartheta}-\eqref{eq:r} to Bloch simulations for the example of anti-periodic boundary conditions.

\section{Efficiency of the Hybrid State}

\begin{figure}[!b]
	\centering
	\begin{tikzpicture}[scale = 1]
	\small 
	\begin{axis}[
	width=\columnwidth*0.8,
	height=\textwidth*0.14,
	scale only axis,
	xmin=-25,
	xmax=85,
	ymin=.625,
	ymax=1,
	xticklabel=\empty,
	ylabel={$T_1$~(s)},
	ylabel style={yshift=-0.1cm},
	clip marker paths=true,
	name=T1
	]
	\addplot [color=UKLblue, mark=o, only marks, error bars/.cd, y dir=both, y explicit,]table[x=sigma, y=T1_mean, y error=T1_std]{Figures/noise_DESPOT.txt};
	\addplot [color=UKLred, mark=x, only marks, error bars/.cd, y dir=both, y explicit,]table[x=sigma, y=T1_mean, y error=T1_std]{Figures/noise_HSFP.txt};
	\addplot [color=black, mark=+, only marks, error bars/.cd, y dir=both, y explicit,]table[x=sigma, y=T1_mean, y error=T1_std]{Figures/noise_MRF.txt};
	
	\addplot[color=black, solid, forget plot] coordinates {(-25, .781) (85, .781)};
	\end{axis}
	
	\begin{axis}[
	width=\columnwidth*0.8,
	height=\textwidth*0.14,
	scale only axis,
	xmin=-25,
	xmax=85,
	ymin=.04,
	ymax=.08,
	xlabel={$\sigma_\omega$~(rad/s)},
	xlabel style={yshift=0.1cm},
	ylabel={$T_2$~(s)},
	ylabel style={yshift=-0.1cm},
	xtick={-20,0,20,40,60,80},
	xticklabels={CRB, $\rightarrow0$, 20,40,60,80},
	scaled y ticks=false,
	yticklabel style={/pgf/number format/fixed},
	clip marker paths=true,
	at=(T1.south east),
	anchor=north east,
	yshift = -0.3cm,
	legend entries = {
		steady state,
		hybrid state,
		transient state
	},
	legend pos = south west,
	]
	\addplot [color=UKLblue, mark=o, only marks, error bars/.cd, y dir=both, y explicit,]table[x=sigma, y=T2_mean, y error=T2_std]{Figures/noise_DESPOT.txt};
	\addplot [color=UKLred, mark=x, only marks, error bars/.cd, y dir=both, y explicit,]table[x=sigma, y=T2_mean, y error=T2_std]{Figures/noise_HSFP.txt};
	\addplot [color=black, mark=+, only marks, error bars/.cd, y dir=both, y explicit,]table[x=sigma, y=T2_mean, y error=T2_std]{Figures/noise_MRF.txt};
	
	\addplot[color=black, solid, forget plot] coordinates {(-25, .065) (85, .065)};
	\end{axis}
	\end{tikzpicture}
	\caption{Both, the steady-state and the hybrid-state experiments are robust with respect to inhomogeneous broadening (here modeled by a Gaussian distribution of Larmor frequencies with the standard deviation $\sigma_\omega$), while the transient state exhibits a substantial bias with increasing broadening. The observed noise (indicated by the error bars) is considerably less in the hybrid state compared to the steady state, and for all experiments the observed noise approximates the limit set by the Cram\'er-Rao bound (CRB) well (far left). 
	The relaxation times were estimated from signal simulated with the steady-state pattern shown in supporting Fig.~\ref{fig:OCT_supp}n, an anti-periodic hybrid-state pattern (Fig.~\ref{fig:OCT_main}e), and the transient state is illustrated using the example of the original magnetic resonance fingerprinting (MRF) \cite{Ma2013} experiment. 
	Note that the steady-state and the hybrid-state experiment have a duration of $T_C = 3.8$~s, while the MRF experiments lasts for $12.3$~s.}
	\label{fig:noise}
\end{figure}
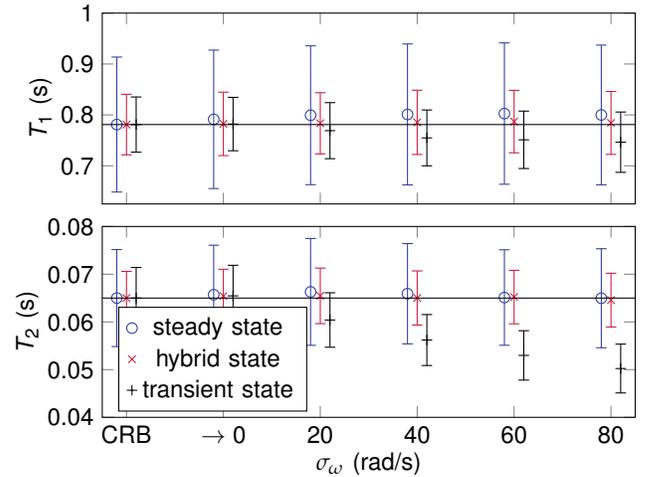

The superior signal-to-noise ratio (SNR) efficiency of the hybrid state in comparison to the steady state becomes evident when comparing numerically optimized experiments. 
For this purpose, we simulated the average signal obtained from a collection of isochromats with a Gaussian distribution of Larmor frequencies and added white noise to reflect thermal noise. Because the internal frequency distribution in a sample is generally unknown, the obtained signals were fitted with their respective models assuming a single isochromat. 
Fig.~\ref{fig:noise} shows that the transient state leads to increasingly biased estimates of the relaxation times as the distribution of Larmor frequencies widens ($\sigma_\omega$ increases). Conversely, both the steady and hybrid state demonstrate a similar robustness with respect to inhomogeneous broadening. As anticipated, the estimates retrieved from the hybrid-state experiment exhibit substantially less noise. The hybrid state, thus, unites superior encoding capabilities similar to the transient state, and robustness deviations of the magnetic fields and to inhomogeneous broadening, similar to the steady state.

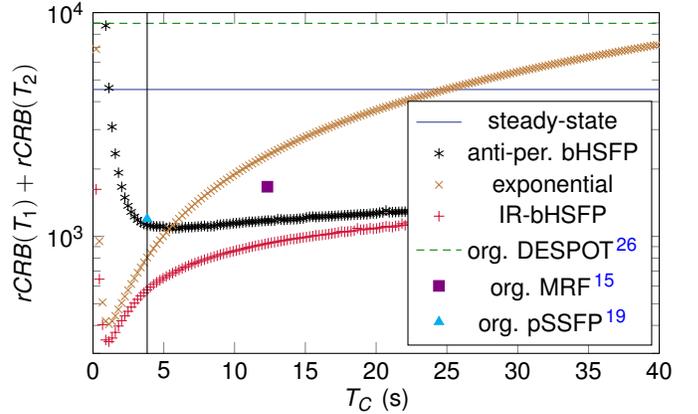
\begin{figure}[tb]
	\centering
	\begin{tikzpicture}[scale = 1]
	\small 
	\begin{semilogyaxis}[
	width=\columnwidth*0.85,
	height=\textwidth*0.25,
	scale only axis,
	xmin=0,
	xmax=40,
	ymin=300,
	ymax=10000,
	xlabel={$T_C~\text{(s)}$},
	xlabel style={yshift=0.1cm},
	ylabel={$rCRB(T_1) + rCRB(T_2)$},
	ylabel style={yshift=-0.1cm},
	legend entries = {
		steady-state,
		anti-per. bHSFP,
		exponential,
		IR-bHSFP,
		org. DESPOT\cite{Deoni2003}, 
		org. MRF\cite{Ma2013},  
		org. pSSFP\cite{Asslander2017}, 
	},
	legend columns = 1,
	legend style={at = {(axis cs:  39.5, 325)}, anchor = south east},
	clip marker paths=true
	]
	\addplot [color=UKLblue,                                     solid]table[x=T_s, y=CRB]{Figures/CRB_T1T2_DESPOT_opt.txt};
	\addplot [color=black, mark=asterisk, only marks]table[x=T_s, y=CRB]{Figures/CRB_T1T2_anti_periodic_pio4.txt};
	\addplot [color=brown, mark=x, only marks]table[x=T_s, y=CRB] {Figures/CRB_T1T2_IR_bSSFP.txt};	
	\addplot [color=UKLred, mark=+, only marks]table[x=T_s, y=CRB]{Figures/CRB_T1T2_r0m1_pio4.txt};

	\addplot [color=green!50!black, densely dashed]table[x=T_s, y=CRB]{Figures/CRB_T1T2_DESPOT_org.txt};
	\begin{pgfonlayer}{foreground}
	\addplot [color=violet, mark=square*, only marks]table[x=T_s, y=CRB]{Figures/CRB_T1T2_orgMRF.txt};
	\addplot [color=cyan, mark=triangle*, only marks]table[x=T_s, y=CRB]{Figures/CRB_T1T2_orgpSSFP.txt};
	\end{pgfonlayer}
			
		
	\addplot[color=black, solid, forget plot] coordinates {(3.825, 50) (3.825, 10000)};
	\end{semilogyaxis}
	\end{tikzpicture}
	\caption{The depicted relative Cram\'er-Rao bounds ($rCRB$) are defined by Eqs.~\eqref{eq:rCRB_T1} and \eqref{eq:rCRB_T2}, and can be understood as a lower bound of the squared inverse SNR efficiency per unit time. One can observe that, for most \textit{cycle times} ($T_C$), exponential decays as well as steady-state experiments are substantially less efficient than variants that exploit the entire experiment design space spanned by the hybrid state, namely the inversion recovery balanced hybrid-state free precession (IR-bHSFP) and the anti-periodic bHSFP experiment. 
	For reference, some experiments from literature are shown as well, namely the original DESPOT \cite{Deoni2003}, MRF \cite{Ma2013},  and pSSFP \cite{Asslander2017} experiment. All Cram\'er-Rao bounds were calculated for the relaxation times $T_1={781}~\text{ms}$ and $T_2={65}~\text{ms}$.}
	\label{fig:rCRB_TC}
\end{figure}

For a more comprehensive analysis of the noise properties of different experiment design spaces, we examine the sum of the relative Cram\'er-Rao bound ($rCRB$) for $T_1$- and $T_2$-encoding. 
The $rCRB$ provides a lower limit for the noise in the estimated parameters, normalized by the input noise variance, by the square of the respective relaxation time and by $T_C/T_R$ (Eqs.~\eqref{eq:rCRB_T1} and \eqref{eq:rCRB_T2}). It can be understood as a lower bound for the squared inverse SNR efficiency per unit time, and Fig.~\ref{fig:noise} shows that the simulated noise comes close to this theoretical limit. 
We numerically searched the parameter space of possible drive functions for the lowest combined $rCRB$. Due to the nature of the steady state, its $rCRB$ does not depend on $T_C$, so that the experiment's duration can be chosen freely to meet the experimental needs. 
Hybrid-state experiments with anti-periodic boundary conditions provide a similar flexibility, since multiple cycles can be concatenated without gaps. Comparing these two experiments, one finds that the hybrid state allows for a substantially more efficient measurement than the steady state (Fig.~\ref{fig:rCRB_TC}). 

\tikzexternaldisable
\tikzexternalenable

The performance of exponential relaxation curves is here demonstrated using the example of the inversion-recovery balanced steady-state free precession (IR-bSSFP) experiment\footnote[2]{Despite the name, this is actually not a steady-state experiment. Instead, one measures the magnetization as it exponentially approaches the steady state.}, which is known to have a high SNR efficiency \cite{Schmitt2004,Ehses2013}. 
In contrast to the previously discussed experiments, the magnetization departs here from thermal equilibrium. This requires a long waiting time ($\Delta t \gg T_1$) before the measurement can be repeated. 
For $T_C \lesssim 25$s, exponential experiments have a lower $rCRB$ compared to steady-state experiments, and for $T_C \lesssim 5$s it is even lower compared to anti-periodic hybrid-state experiments (Fig.~\ref{fig:rCRB_TC}). 
An optimization of exponential experiments is essentially the search for the optimal line from the southern half of the Bloch sphere to the steady-state ellipse (supporting Fig.~\ref{fig:OCT_supp}g). 
If we take the IR-bSSFP experiment and allow $\vartheta(t)$ to vary over time, we can exploit the full experiment design space spanned by the hybrid state, and we find an improved SNR-efficiency at all $T_C$ values, with the most dramatic improvement in the case of long experiments. In analogy to the acronym IR-bSSFP, we use the term inversion-recovery balanced hybrid-state free precession (IR-bHSFP) for hybrid-state experiments that start from thermal equilibrium by the application of an inversion pulse\footnote[3]{We focus this analysis on experiments with balanced gradient moments because of their superior SNR properties.}. 

In this section, we analyzed the noise properties at a single $T_1$ and $T_2$ value. Supporting Figs.~\ref{fig:rCRB_T1_T2} and \ref{fig:rCRB_B0_B1} demonstrate that the conclusions drawn here remain valid throughout large areas in $T_1$-$T_2$-space, and also in the presence of deviations of the Larmor and Rabi frequencies. 


\section{Spin Dynamics in the Hybrid State}
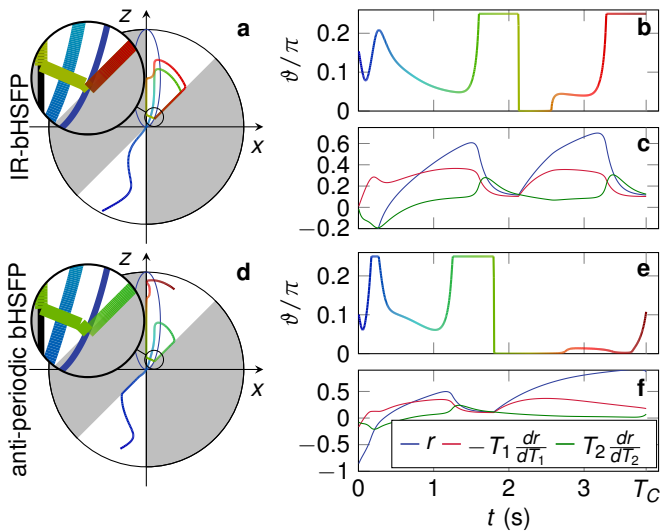
\begin{figure}[tb]
	\centering
	\begin{tikzpicture}[scale = 1]
	\def\TR{0.0045};
	\def\TOne{0.781}; 
	\def\TTwo{0.065};
	\def\Theta{22.5};
	
	\small 
	\begin{scope}[spy using outlines={circle, magnification=6, connect spies}]
	\begin{axis}[
	width=\textwidth*0.3,
	axis equal image, 
	axis lines=center, 
	xmin = -1.2, xmax = 1.2,
	ymin = -1.2, ymax = 1.2,
	xtick=\empty, ytick=\empty, ztick=\empty,
	xlabel={$x$}, ylabel={$z$}, 
	every axis x label/.style={at={(ticklabel cs:0.975)}, anchor=north},
	every axis y label/.style={at={(ticklabel cs:0.975)}, anchor=east},
	colormap name = mybluered, 
	point meta min=0,
	point meta max= 3.825,
	name=Bloch_IR,
	clip=false,
	]
	
	\begin{pgfonlayer}{background}
	\filldraw[fill=lightgray] (axis cs: 0,0) -- (axis cs: 0, 1) arc [start angle=  90, end angle= 225, radius={transformdirectionx(1)}];
	\filldraw[fill=lightgray] (axis cs: 0,0) -- (axis cs: 0,-1) arc [start angle=270, end angle=405, radius={transformdirectionx(1)}];
	\end{pgfonlayer}
	
	\addplot[no marks, domain=0:360,samples=60,] ({sin(x)}, {cos(x)});
	
	\addplot[domain=0:1,samples=1000, UKLblue] ({-sqrt(\TTwo * (1/(4*\TOne) - (x - .5)^2/\TOne)}, {x});
	\addplot[domain=0:1,samples=1000, UKLblue] ({  sqrt(\TTwo * (1/(4*\TOne) - (x - .5)^2/\TOne)}, {x});
	
	\addplot[mesh, point meta=explicit, thick, line join=bevel]table[x=y, y=z, meta=t_s]{Figures/OCT_T1T2_r0m1_pio4.txt};
	
	\coordinate (spypoint) at (axis cs:0.08,0.1);
	\coordinate (spyviewer) at (axis cs:-0.6,0.5);
	\spy[black,size=1.5cm] on (spypoint) in node (myspy) [fill=white] at (spyviewer);
	
	\node[above, inner sep=0mm,minimum size=5mm, rotate=90] at (axis cs:  -1.1,0) {IR-bHSFP};
	\node[right, inner sep=0mm,minimum size=5mm] at (axis cs:  .8,1) {\textbf{a}};
	\end{axis}
	\end{scope}
	
	\begin{axis}[
	width=\textwidth*0.22,
	height=\textheight*0.055,
	scale only axis,
	xmin=0,
	xmax=3.99,
	ymin=0,
	ymax=.26,
	extra x ticks={3.825}, extra x tick labels={},
	xticklabel=\empty,
	ylabel={$\vartheta/\pi$},
	colormap name = mybluered, 
	point meta min=0,
	point meta max= 3.825,
	name=theta_IR,
	at=(Bloch_IR.right of north east),
	anchor=left of north west,
	]
	\addplot [mesh, thick, point meta=explicit, solid]table[x=t_s, y=theta, meta=t_s]{Figures/OCT_T1T2_r0m1_pio4.txt};
	
	\node[right, inner sep=0mm] at (axis cs:  3.7, {(.15+.2)/.42*.26}) {\textbf{b}};
	\end{axis}	
	
	\begin{axis}[
	width=\textwidth*0.22,
	height=\textheight*0.055,
	scale only axis,
	xmin=0,
	xmax=3.99,
	ymin=-.2,
	ymax=.75,
	extra x ticks={3.825}, extra x tick labels={},
	xticklabel=\empty,
	name=Fisher_IR,
	at=(theta_IR.below south east),
	anchor= north east,
	trig format plots=rad,
	]
	\addplot [color=UKLblue,            solid]table[x=t_s, y=r        ]{Figures/OCT_T1T2_r0m1_pio4.txt};
\addplot [color=UKLred,              solid]table[x=t_s, y=drdT1]{Figures/OCT_T1T2_r0m1_pio4.txt};
\addplot [color=green!50!black, solid]table[x=t_s, y=drdT2]{Figures/OCT_T1T2_r0m1_pio4.txt};
	
	\node[right, inner sep=0mm] at (axis cs:  3.7, {(.15+.2)/.42*.95-.2}) {\textbf{c}};
	\end{axis}
		
	\begin{scope}[spy using outlines={circle, magnification=6, connect spies}]
	\begin{axis}[
	width=\textwidth*0.3,
	axis equal image, 
	axis lines=center, 
	xmin = -1.2, xmax = 1.2,
	ymin = -1.2, ymax = 1.2,
	xtick=\empty, ytick=\empty, ztick=\empty,
	xlabel={$x$}, ylabel={$z$}, 
	every axis x label/.style={at={(ticklabel cs:0.975)}, anchor=north},
	every axis y label/.style={at={(ticklabel cs:0.975)}, anchor=east},
	colormap name = mybluered, 
	point meta min=0,
	point meta max= 3.825,
	name=Bloch_anti,
	at=(Bloch_IR.below south west),
	anchor=above north west,
	clip=false,
	]
	
	\begin{pgfonlayer}{background}
	\filldraw[fill=lightgray] (axis cs: 0,0) -- (axis cs: 0, 1) arc [start angle=  90, end angle= 225, radius={transformdirectionx(1)}];
	\filldraw[fill=lightgray] (axis cs: 0,0) -- (axis cs: 0,-1) arc [start angle=270, end angle=405, radius={transformdirectionx(1)}];
	\end{pgfonlayer}
	
	\addplot[no marks, domain=0:360,samples=60,] ({sin(x)}, {cos(x)});
	
	\addplot[domain=0:1,samples=1000, UKLblue] ({-sqrt(\TTwo * (1/(4*\TOne) - (x - .5)^2/\TOne)}, {x});
	\addplot[domain=0:1,samples=1000, UKLblue] ({  sqrt(\TTwo * (1/(4*\TOne) - (x - .5)^2/\TOne)}, {x});
	
	\addplot[mesh, point meta=explicit, thick, line join=bevel]table[x=y, y=z, meta=t_s]{Figures/OCT_T1T2_anti_periodic_pio4.txt};
	
	\coordinate (spypoint) at (axis cs:0.08,0.1);
	\coordinate (spyviewer) at (axis cs:-0.6,0.5);
	\spy[black,size=1.5cm] on (spypoint) in node (myspy) [fill=white] at (spyviewer);
	
	\node[above, inner sep=0mm,minimum size=5mm, rotate=90] at (axis cs:  -1.1,0) {anti-periodic bHSFP};
	\node[right, inner sep=0mm,minimum size=5mm] at (axis cs:  .8,1) {\textbf{d}};
	\end{axis}
	\end{scope}
	
	\begin{axis}[
	width=\textwidth*0.22,
	height=\textheight*0.055,
	scale only axis,
	xmin=0,
	xmax=3.99,
	ymin=0,
	ymax=.26,
	extra x ticks={3.825}, extra x tick labels={},
	xticklabel=\empty,
	ylabel={$\vartheta/\pi$},
	colormap name = mybluered, 
	point meta min=0,
	point meta max= 3.825,
	name=theta_anti,
	at=(Bloch_anti.right of north east),
	anchor=left of north west,
	]
	\addplot [mesh, thick, point meta=explicit, solid]table[x=t_s, y=theta, meta=t_s] {Figures/OCT_T1T2_anti_periodic_pio4.txt};
	
	\node[right, inner sep=0mm] at (axis cs:  3.7, {(.15+.2)/.42*.26}) {\textbf{e}};
	\end{axis}	
	
	\begin{scope}[spy using outlines={magnification=3, connect spies}]
	\begin{axis}[
	width=\textwidth*0.22,
	height=\textheight*0.055,
	scale only axis,
	xmin=0,
	xmax=3.99,
	ymin=-1,
	ymax= .9,
	xlabel={$t~\text{(s)}$},
	extra x ticks={3.825}, extra x tick labels={$T_C$},
	xlabel style={yshift=0.15cm},
	name=Fisher_anti,
	at=(theta_anti.below south east),
	anchor= north east,
	legend entries = {$r$, $-T_1\frac{dr}{dT_1}$, $T_2\frac{dr}{dT_2}$},
	legend pos = south east,
	legend columns=3,
	legend style={
		xshift = 0.1cm,
		legend image code/.code={\draw[##1,line width=0.6pt] plot coordinates {(0cm,0cm) (0.25cm,0cm)};}
	},
	]
	\addplot [color=UKLblue,            solid]table[x=t_s, y=r        ]{Figures/OCT_T1T2_anti_periodic_pio4.txt};
	\addplot [color=UKLred,              solid]table[x=t_s, y=drdT1]{Figures/OCT_T1T2_anti_periodic_pio4.txt};
	\addplot [color=green!50!black, solid]table[x=t_s, y=drdT2]{Figures/OCT_T1T2_anti_periodic_pio4.txt};

	\node[right, inner sep=0mm] at (axis cs:  3.7,  {(.15+.2)/.42*1.9-1}) {\textbf{f}};
	
	\end{axis}
	\end{scope}
	\end{tikzpicture}
	\caption{The spin dynamics in hybrid state experiments are depicted on Bloch spheres (\textbf{a},\textbf{d}). The optimized polar angle functions are shown in (\textbf{b},\textbf{e}), with the color scale providing a reference for the trajectories on the Bloch spheres. The radial component magnetization and its normalized derivatives with respect to the relaxation times are the foundation of computing the relative Cram\'er-Rao bound and are shown in (\textbf{c},\textbf{f}). Both spin trajectories were jointly optimized for $T_1$ and $T_2$ and the polar angle was limited to $0 \leq \vartheta \leq \pi/4$.
	}
	\label{fig:OCT_main}
\end{figure}

Optimizing the driving functions $\vartheta(t)$ results in spin trajectories with reproducible features.
For example, all optimizations resulted in comparatively smooth functions $\vartheta(t)$. Note that the optimizations assume a hybrid state, but otherwise do not enforce smoothness, which indicates that the adiabaticity condition (Eq.~\eqref{eq:adiabaticity_HSFP}) does not impair the $T_{1,2}$-encoding efficiency. 
In some segments, the optimization exploits the design limits $0 \leq \vartheta \leq \pi/4$, which are imposed for practical reasons. These extreme values help to achieve a large $dr/dT_1$ while minimizing $dr/dT_2$ and vice versa. However, in other segments, e.g., directly after crossing the origin (turquoise segment), the derivative $dr/dT_2$ is already close to zero and the magnetization follows a trajectory with $\vartheta > 0$. Similarly, after a segment of $\vartheta \approx 0$ (yellow segment), $dr/dT_2$  approaches zero and the optimized driving function transitions to a $\vartheta > 0$, resulting in non-zero signal and disentangled encoding of $r$ and $dr/dT_1$. 
Further, the optimized trajectories do not spend a significant amount of time on the steady-state ellipse. On the contrary, crossing the ellipse triggers a fast change of $\vartheta$, as highlighted by the magnifications in Fig.~\ref{fig:OCT_main}. 

Described hybrid-state spin trajectories result from non-convex optimizations and we can only speculate about their optimality. However, the simple and reproducible structures, together with the simple form of the governing Eq.~\eqref{eq:Bloch_2D} provide an excellent basis for a more detailed analysis.

\section{In Vivo Experiment}
\begin{figure}[t]
	\centering
	\begin{tikzpicture}[scale = 0.775]
	\begin{axis}[%
	width={6cm/256*200},
	height=6cm,
	axis on top,
	scale only axis,
	xmin=0,
	xmax=1,
	y dir=reverse,
	ymin=0,
	ymax=1,
	hide axis,
	name=T1_cTSFP_cartesian,
	align=center,
	title={Bloch model},
	title style = {yshift=-.2cm},
	]
	\addplot graphics [xmin=0,xmax=1,ymin=0,ymax=1] {In_vivo_sag/T1_cTSFP_Bloch_slice_69_org.png};
	
	\draw [UKLred, thick](axis cs: {96/200}, {107/256}) rectangle (axis cs: {101/200}, {111/256});
	\end{axis}
	
	\begin{axis}[%
	width={6cm/256*200},
	height=6cm,
	axis on top,
	scale only axis,
	xmin=0,
	xmax=1,
	y dir=reverse,
	ymin=0,
	ymax=1,
	hide axis,
	name=T2_cTSFP_cartesian,
	at=(T1_cTSFP_cartesian.below south west),
	anchor=above north west,
	yshift=-.08in,
	]
	\addplot graphics [xmin=0,xmax=1,ymin=0,ymax=1] {In_vivo_sag/T2_cTSFP_Bloch_slice_69_org.png};
	\end{axis}
	
	\begin{axis}[%
	width={6cm/256*200},
	height=6cm,
	axis on top,
	scale only axis,
	xmin=0,
	xmax=1,
	y dir=reverse,
	ymin=0,
	ymax=1,
	hide axis,
	name=T1_cTSFP_radial,
	at=(T1_cTSFP_cartesian.right of north east),
	anchor= left of north west,
	xshift= .08in,
	align=center,
	title={hybrid-state model},
	title style = {yshift=-.2cm},
	colormap name = parula,
	colorbar,
	point meta min=0.668,
	point meta max=5,
	colorbar style={ymode=log, ytick={1,2,3,4,5}, ylabel=$T_1~(\text{s})$, log ticks with fixed point, ylabel style = {yshift = 0cm}, width=0.4cm, xshift=-0.1cm},
	]
	\addplot graphics [xmin=0,xmax=1,ymin=0,ymax=1] {In_vivo_sag/T1_cTSFP_radial_slice_69_org.png};
	\end{axis}
	
	\begin{axis}[%
	width={6cm/256*200},
	height=6cm,
	axis on top,
	scale only axis,
	xmin=0,
	xmax=1,
	y dir=reverse,
	ymin=0,
	ymax=1,
	hide axis,
	name=T2_cTSFP_radial,
	at=(T1_cTSFP_radial.below south west),
	anchor=above north west,
	yshift=-.08in,
	colormap name = parula,
	colorbar,
	point meta min=0.0316,
	point meta max=0.316,
	colorbar style={ymode=log, ytick={0.05,0.1,0.2,0.3}, ylabel=$T_2~(\text{s})$, log ticks with fixed point, ylabel style = {yshift = 0.3cm}, width=0.4cm, xshift=-0.1cm},
	]
	\addplot graphics [xmin=0,xmax=1,ymin=0,ymax=1] {In_vivo_sag/T2_cTSFP_radial_slice_69_org.png};
	\end{axis}
	
	
	\end{tikzpicture}
	\caption{A single sagittal slice of an in vivo 3D human brain MRI scan is depicted. The data were acquired with an anti-periodic bHSFP experiment and were fitted once with the Bloch model (Eq.~\eqref{eq:Bloch}), and once with the hybrid-state model (Eq.~\eqref{eq:vartheta}-\eqref{eq:r}). The parameter maps have a resolution of $1~\textnormal{mm} \times 1~\textnormal{mm} \times 2~\textnormal{mm}$ and spatial encoding was performed with a 3D stack-of-stars k-space trajectory\cite{Chandarana2011}. The red box indicates a region of interest used for extracting $T_1$ and $T_2$ values. Note the logarithmic scale of the color coding. The entire 3D data set can be found in supporting Fig.~\ref{fig:InVivo_sag_all}.}
	\label{fig:InVivo_sag}
\end{figure}
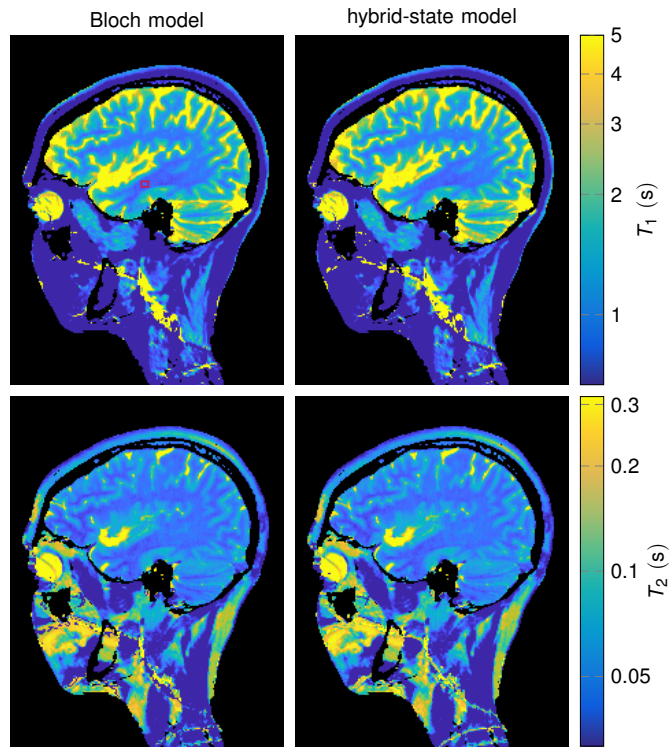

Fig.~\ref{fig:InVivo_sag} shows an example application of the hybrid state. The $T_1$- and $T_2$-maps in a sagittal slice through a human brain were acquired with an anti-periodic bHSFP experiment and also serve as a validation of the hybrid-state model: Fitting the data with the full Bloch model and the hybrid-state model resulted in virtually the same $T_1$- and $T_2$-maps, which is also confirmed by the values within a region of interest (Bloch model: $T_1 = 965 \pm 23$ms, $T_2 = 48.2 \pm 3.0$ms; hybrid-state model: $T_1 = 988 \pm 23$ms, $T_2 = 49.7 \pm 2.9$ms). 

\section{Scope of the Hybrid-State Model}
Adiabatic passages are frequently used in NMR, MRI, as well as quantum computing for robust spin excitation, inversion, and refocusing in the presence of magnetic field inhomogeneities \cite{Silver1984,Jones2000}. These passages are achieved by continuous, slowly varying driving fields, and are commonly assumed to be much faster than spin relaxation, such that one enforces adiabatic transitions of the magnetization's direction, while its magnitude is assumed to be constant. Neglecting relaxation, the Hamiltonian in Eq.~\eqref{eq:Bloch} reduces to a generator of a rotation and we can derive the well established adiabaticity condition $|d\omega_{x,y,z}/dt| \ll {\omega_{x}^2 + \omega_{y}^2 + \omega_{z}^2}$ with the described formalism. 
Here, we generalized adiabatic passages to pulsed experiments, which allows for exploiting their robustness throughout the entire experiment. 
The hybrid-state adiabaticity condition (Eq.~\eqref{eq:adiabaticity_HSFP}) has a very similar structure to the established adiabaticity condition, apart from an additional relaxation term, which is required at typical experiment durations at the order of seconds to minutes. 
Gaining a flexible and efficient access to relaxation mechanisms while exploiting the robustness of adiabatic passages constitutes the core of the hybrid-state framework.

The robustness of the measured signal to magnetic field deviations, including inhomogeneous broadening, is reflected by the hybrid-state equations of motion (Eqs.~\eqref{eq:vartheta}-\eqref{eq:r}) being smooth functions of the Larmor and Rabi frequencies, which are here parameterized by $\phi$ and $\alpha$, respectively. This property is a direct consequence of constraining the population of the complex eigenstates and is particularly important when the line shape is unknown, e.g., when measuring biological tissue with balanced-HSFP experiments \cite{Ganter2006}. 
The estimation of the distribution is less problematic in unbalanced experiments, such as the fast imaging with steady-state precession \cite{Oppelt1986} (FISP) experiment, or the reversed PSIF experiment. In these experiments, one places spoiler gradient pulses directly before or after the RF pulses, which 
desensitize the signal to inhomogeneous broadening at the cost of SNR. The hybrid-state model holds true for these experiments, and the spoiler gradients can be incorporated by setting $\phi_{T_E} = 0$ or $\phi_{T_E} = \phi$ in Eq.~\eqref{eq:varphi} 
for FISP and PSIF, respectively. 



For complex molecules, as well as for complex biological tissues, the Bloch equation is an oversimplified model. This can be observed in Fig.~\ref{fig:InVivo_sag}, where the measured relaxation times are subject to systematic deviations, which are most likely caused by magnetization transfer \cite{Wolff1989,Bieri2006,Hilbert2016}. Magnetization transfer, as well as diffusion \cite{Torrey1956} and chemical exchange \cite{McConnell1958}, are captured neither by the Bloch equation, nor by the hybrid-state model in their basic form. However, these effects can be modeled by extensions to the hybrid-state model similarly to the established extensions of the Bloch equation \cite{Torrey1956,McConnell1958}. Such extended hybrid-state models can provide a more intuitive understanding of these effects, and pave the road towards more efficient experiment designs to measure them.

\newpage
\small
\section{Methods} \label{Methods}
\subsection{Adiabaticity Conditions of the Hybrid State} \label{sec:Adiabaticity}
\subsubsection{The Evolution Matrix}
In order to describe pulsed MR experiments, we analyze the spin evolution matrix $\mathbf{U} \in \mathbb{R}^{4 \times 4}$, which is generated by the Hamiltonian. The matrix $\mathbf{U}$ can, e.g., be derived by taking the matrix exponential of the Hamiltonian and is not unitary due to the relaxation terms (Eq.~\eqref{eq:Bloch}). Note that an analysis of the evolution matrix is largely equivalent to an analysis based on the Hamiltonian itself. For pulsed experiments, where we assume one hard, i.e. infinitesimally short, RF pulse, surrounded by Larmor precession and relaxation, the evolution matrix is given by 
\begin{equation}
\mathbf{U} = \mathbf{E} \cdot \mathbf{R}_z \cdot \mathbf{R}_y \cdot \mathbf{R}_z \cdot \mathbf{E},
\label{eq:U_def}
\end{equation}
where
\begin{equation*}
\mathbf{E} = \begin{pmatrix}
\sqrt{E_2} &                0 &               0 & 0 \\
0 & \sqrt{E_2} &               0 & 0 \\
0 &                0 & \sqrt{E_1} & 1-\sqrt{E_1} \\
0 &                0 &               0 & 1 \\
\end{pmatrix}
\end{equation*}
describes the relaxation of the magnetization with $E_{1,2} = \exp (-T_R/T_{1,2})$. The rotation matrices
\begin{equation*}
\mathbf{R}_y = \begin{pmatrix}
\cos \alpha & 0 & -\sin \alpha & 0 \\
0 & 1 & 0 & 0 \\
\sin \alpha & 0 & \cos \alpha  & 0 \\
0 & 0 &                   0 & 1 \\
\end{pmatrix}
\end{equation*}
and
\begin{equation*}
\mathbf{R}_z = \begin{pmatrix}
\cos \frac{\phi}{2} & -\sin \frac{\phi}{2} & 0 & 0 \\
\sin \frac{\phi}{2}  &  \cos \frac{\phi}{2} & 0 & 0 \\
0 &                  0 &  1 &  0 \\
0 &                  0 &  0 & 1 \\
\end{pmatrix} 
\end{equation*}
describe the rotations caused by the RF pulse and free precession, respectively. \footnote[5]{Eq. \eqref{eq:U_def} assumes a symmetric experiment, as it is used e.g. in balanced-SSFP experiments, where one usually measures the magnetization in the middle between two RF pulses ($T_E = T_R/2$) \cite{Scheffler2003}. In the case of unbalanced-SSFP experiments, one would usually acquire the magnetization right after each RF pulse and would place a so-called spoiler gradient after the signal acquisition in order to create a net gradient moment. In such a FISP \cite{Oppelt1986} experiment, the evolution matrix would, thus, be given by $\mathbf{U}_{\text{FISP}} = \mathbf{R}_y \cdot \mathbf{R}_z \cdot \mathbf{E}^2$ with the appropriate choice of $\phi$, and the reversed PSIF experiment with the spoiler gradient prior to the readout would be described by $\mathbf{U}_{\text{PSIF}} = \mathbf{E}^2 \cdot \mathbf{R}_z \cdot \mathbf{R}_y$. Note that derivations for FISP and PSIF lead to the same result as the one presented here. }

For future reference, we also define the derivative  of $\mathbf{U}$ with respect to $\alpha$, which is given by $\mathbf{U}' = \mathbf{E} \mathbf{R}_z \mathbf{R}_y' \mathbf{R}_z \mathbf{E}$ with
\begin{equation}
\mathbf{R}_y' = \begin{pmatrix}
-\sin \alpha & 0 &  -\cos \alpha & 0 \\
0 & 0 & 0 & 0 \\
\cos \alpha & 0 & -\sin \alpha  & 0 \\
0 & 0 &                   0 & 0 \\
\end{pmatrix},
\label{eq:Ry'}
\end{equation}
and the derivative of $\mathbf{U}$ with respect to $\phi$, which is given by $\mathbf{U}' = \mathbf{E} \mathbf{R}_z' \mathbf{R}_y \mathbf{R}_z \mathbf{E} + \mathbf{E} \mathbf{R}_z \mathbf{R}_y \mathbf{R}_z' \mathbf{E}$ with

\begin{equation}
\mathbf{R}_z' = \frac{1}{2} \begin{pmatrix}
-\sin \frac{\phi}{2} & -\cos \frac{\phi}{2} & 0 & 0 \\
\cos \frac{\phi}{2}  &  -\sin \frac{\phi}{2} & 0 & 0 \\
0 &                  0 &  0 &  0 \\
0 &                  0 &  0 &  0 \\
\end{pmatrix}.
\label{eq:Rz'}
\end{equation}

\vspace{0.15cm}
\subsubsection{Eigendecomposition of the Evolution Matrix}
The eigendecomposition of the evolution matrix is given by 
\begin{equation}
	\mathbf{U} = \mathbf{V} \boldsymbol{\Lambda} \mathbf{V}^{-1},
\end{equation}
where $\mathbf{V} \in \mathbb{C}^{4 \times 4}$ is composed of the right-eigenvectors $\mathbf{v}_d \in \mathbb{C}^{4 \times 1}$ defined by $\mathbf{U}\mathbf{v}_d = \lambda_d \mathbf{v}_d$, and $\boldsymbol{\Lambda} \in \mathbb{C}^{4 \times 4}$ is a diagonal matrix with the eigenvalues $\lambda_d \in \mathbb{C}$ on the diagonal. The magnetization in MR experiments never grows arbitrarily, so that $|\lambda_d| \leq 1$ must be fulfilled for all eigenvalues. Further, if the experiment described by $\mathbf{U}$ has a non-zero steady-state magnetization, at least one eigenvalue must fulfill $|\lambda_d| = 1$. 

For the explicit definition of the evolution matrix in Eq.~\eqref{eq:U_def}, which describes one RF pulse surrounded by free precession and relaxation, one eigenvalue is given by
\begin{equation}
	\lambda_{\text{S}} = 1
\end{equation}
and the corresponding eigenvector describes the steady-state magnetization. As shown by Ganter \cite{Ganter2004}, the remaining eigenvalues are approximated by 
\begin{align}
\lambda_\parallel &= \frac{1}{\eta^2} \left(\cos^2 \frac{\alpha}{2} \sin^2 \frac{\phi}{2} E_1 + \sin^2 \frac{\alpha}{2} E_2 \right) 
\label{eq:lambda_par}\\
\lambda_\bot^{(*)} &= \frac{e^{\pm i \Omega}}{2\eta^2} \left(\sin^2 \frac{\alpha}{2} E_1 + \left(\eta^2 +  \cos^2 \frac{\alpha}{2} \sin^2 \frac{\phi}{2} \right) E_2 \right)
\label{eq:lambda_orth}
\end{align}
with 
\begin{align}
\eta &= \sqrt{\cos^2 \frac{\alpha}{2} \sin^2 \frac{\phi}{2} + \sin^2 \frac{\alpha}{2}}\\
e^{\pm i \Omega} &= 1 - 2 \eta^2 \pm 2 \eta i \cos \frac{\alpha}{2} \cos \frac{\phi}{2}.
\label{eq:lambda_orth_phase}
\end{align}
These eigenvalues are a first order approximation of the parameter 
\begin{equation}
	\delta = \frac{E_1 - E_2}{E_1 + E_2},
\end{equation}
which is small for $T_R \ll \{T_1,T_2\}$ in most biological tissues \cite{Ganter2004}, have an absolute value smaller than one, and describe the transient state. The eigenvalue $\lambda_\parallel$ is real-valued and the corresponding eigenvector is approximately parallel to the steady-state magnetization in the three spatial dimensions \cite{Ganter2004}. The other two eigenvalues $\lambda_\bot^{(*)}$ are in general complex and complex conjugate of each other, as indicated by the star. This results in the well known oscillatory behavior of the transient state of bSSFP experiments\cite{Hargreaves2001}. As shown by Ganter \cite{Ganter2004}, the corresponding eigenvectors are approximately perpendicular to the steady-state eigenvector. 

\vspace{0.15cm}
\subsubsection{The Perturbation Matrix}
A sequence of $N$ identical and equidistant RF pulses is simply described by $\mathbf{U}^N = \mathbf{V} \boldsymbol{\Lambda}^N \mathbf{V}^{-1}$ and describes the transition into the steady state \cite{Hargreaves2001,Ganter2004}. The description of an experiment with varying driving fields, as required to avoid the steady state, is slightly more complicated. To approach this problem, we denote the evolution matrix of the $n^\text{th}$ repetition by $\mathbf{U}_n$ and the spin dynamics in two consecutive repetitions is described by $\mathbf{U}_n \mathbf{U}_{n-1} = \mathbf{V}_n \boldsymbol{\Lambda}_n \mathbf{V}_n^{-1} \mathbf{V}_{n-1} \boldsymbol{\Lambda}_{n-1} \mathbf{V}_{n-1}^{-1}= \mathbf{V}_n \boldsymbol{\Lambda}_n \mathbf{P}_{n} \boldsymbol{\Lambda}_{n-1} \mathbf{V}_{n-1}^{-1}$. Here, the perturbation matrix 
\begin{equation}
\mathbf{P}_{n} = \mathbf{V}_n^{-1} \mathbf{V}_{n-1}
\label{eq:P_def}
\end{equation}
describes the transformation from the eigenspace of $\mathbf{U}_{n-1}$ to the eigenspace of $\mathbf{U}_n$. 

\vspace{0.15cm}
\subsubsection{Expanding the Perturbation Matrix}
Since an explicit notation of the perturbation matrix is not very enlightening, we approximate its elements by a Taylor expansion. As demonstrated in the supporting material, any changes $\Delta \kappa$ of the parameters $\kappa \in \{\alpha, \phi\}$ has to be small in order to avoid a population of the transient eigenstates. This allows us to employ the Taylor expansion $\mathbf{U}_{n-1} = \mathbf{U}(\kappa_{n-1}) = \mathbf{U}(\kappa_n) - \Delta \kappa_n \mathbf{U}'(\kappa_n) + \mathcal{O}(\Delta \kappa_n^2)$, where $\mathbf{U}'(\kappa_n) = d\mathbf{U}/d \kappa ~|_{\kappa = \kappa_n}$ denotes the derivative evaluated at $\kappa_n$. Assuming that $\mathbf{U}(\kappa_n)$ is not degenerate, i.e. all eigenvalues are distinct, we can utilize the Taylor series described by Eq.~(10.2) in Chapter~2 of Ref. \cite{Wilkinson1965} to expand the perturbation matrix (Eq. \eqref{eq:P_def}). The diagonal elements are then given by $P_{d \rightarrow d} = 1$ and the off-diagonal elements by
\begin{equation}
P_{d \rightarrow f \neq d}(\kappa_n, \Delta \kappa_n) \approx \frac{\Delta \kappa_n \;\; \mathbf{u}_f^H(\kappa_n) \mathbf{U}'(\kappa_n) \mathbf{v}_d(\kappa_n)}{(\lambda_d(\kappa_n) - \lambda_f(\kappa_n)) \mathbf{u}_f^H(\kappa_n) \mathbf{v}_f(\kappa_n)},
\label{eq:P_elements}
\end{equation}
where the left-eigenvectors are defined by $\mathbf{u}_f^H(\kappa_n) \mathbf{U}(\kappa_n) = \lambda_f(\kappa_n) \mathbf{u}_f^H(\kappa_n)$ and the right-eigenvectors by $\mathbf{U}(\kappa_n) \mathbf{v}_f(\kappa_n) = \lambda_f(\kappa_n) \mathbf{v}_f(\kappa_n)$. The superscript $H$ indicates the complex conjugate transpose. Eq.~\eqref{eq:P_elements} has some similarities to the quantum mechanical adiabatic theorem \cite{Born1928}. In both cases, the matrix elements strongly depend on the gap between the eigenvalues. Like in quantum mechanical case, $\lambda_S - \lambda_\parallel$ is purely determined by the absolute value of the eigenvalues, since they both are real-valued and positive. This is fundamentally different in the case of $\lambda_S - \lambda_\bot^{(*)}$, where the gap is dominated by the complex phase of $\lambda_\bot^{(*)}$. In the following, we will show that this key difference opens the door for the hybrid state to emerge. 

\vspace{0.15cm}
\subsubsection{The Population of the Transient Eigenstates}
In order to analyze the cumulative population transfer during $N$ repetitions, we describe the corresponding spin dynamics by 
\begin{equation}
\prod_{n=1}^{N} \mathbf{U}_{N-n} = \mathbf{V}_{N-1} \left(\prod_{n=1}^{N-1} \boldsymbol{\Lambda}_{N-n} \mathbf{P}_{N-n} \right) \boldsymbol{\Lambda}_0 \mathbf{V}_0^{-1}.
\label{eq:N_RF_Pulses}
\end{equation}
The goal of this section is to extract the essential elements of this matrix product and to derive boundary conditions for avoiding a population of the individual eigenstates that describe the transient state magnetization. For this purpose, we will first show that only the population transfer from the steady state is of relevance.

The steady-state left-eigenvector $\mathbf{u}_{\text{S}}^H = (0, 0, 0, 1)$ becomes evident by multiplying it from the left to $\mathbf{U}$ (Eq. \eqref{eq:U_def}). For either parameter variation, we obtain $\mathbf{u}_{\text{S}}^H\mathbf{U}' = (0,0,0,0)$ since the last rows of $\mathbf{R}_y'$ and $\mathbf{R}_z'$ contain only zeros (Eqs.~\eqref{eq:Ry'}, \eqref{eq:Rz'}). With Eq.~\eqref{eq:P_elements}, it follows that $P_{d \rightarrow S} = 0 \forall d \neq S$, resulting in the following structure of the perturbation matrix: 
\begin{equation*}
\begin{split}
&\mathbf{P}_n \approx \\ &\begin{pmatrix}
1 & 0 & 0 & 0 \\
P_{\text{S} \rightarrow \parallel}(\kappa_n,\Delta \kappa_n) + \mathcal{O}(\Delta \kappa_n^2) & 1 & \mathcal{O}(\Delta \kappa_n) & \mathcal{O}(\Delta \kappa_n) \\
P_{\text{S} \rightarrow \bot}(\kappa_n,\Delta \kappa_n) + \mathcal{O}(\Delta \kappa_n^2) & \mathcal{O}(\Delta \kappa_n)  & 1 & \mathcal{O}(\Delta \kappa_n) \\
P_{\text{S} \rightarrow \bot}^*(\kappa_n,\Delta \kappa_n) + \mathcal{O}(\Delta \kappa_n^2) & \mathcal{O}(\Delta \kappa_n) & \mathcal{O}(\Delta \kappa_n) & 1 \\
\end{pmatrix}
\end{split}
\end{equation*}
Here, only the essential elements are denoted explicitly. The central part of Eq. \eqref{eq:N_RF_Pulses} describes the combined effect of $N$ RF pulses with varying parameters onto the eigenvectors and is given by
\begin{widetext}
	\begin{equation}
	\prod_{n=0}^{N-1} \boldsymbol{\Lambda}_{N-n} \mathbf{P}_{N-n} \approx
	\begin{pmatrix}
	1 & 0 & 0 & 0 \\
	\sum_{n=1}^{N} P_{\text{S} \rightarrow \parallel}(\kappa_n,\Delta \kappa_n) \prod_{k=n}^{N} \lambda_\parallel(\kappa_k) + \mathcal{O}(\Delta \kappa_n^2) &  \mathcal{O}(\lambda^N) & \mathcal{O}(\Delta \kappa \cdot \lambda^N) & \mathcal{O}(\Delta \kappa \cdot \lambda^N) \\
	\sum_{n=1}^{N} P_{\text{S} \rightarrow \bot}(\kappa_n,\Delta \kappa_n) \prod_{k=n}^{N} \lambda_\bot(\kappa_k) + \mathcal{O}(\Delta \kappa_n^2) &  \mathcal{O}(\Delta \kappa \cdot \lambda^N) & \mathcal{O}(\lambda^N) & \mathcal{O}(\Delta \kappa \cdot \lambda^N) \\
	\sum_{n=1}^{N} P_{\text{S} \rightarrow \bot}^*(\kappa_n,\Delta \kappa_n) \prod_{k=n}^{N} \lambda_\bot^*(\kappa_k) + \mathcal{O}(\Delta \kappa_n^2) &  \mathcal{O}(\Delta \kappa \cdot \lambda^N) & \mathcal{O}(\Delta \kappa \cdot \lambda^N) & \mathcal{O}(\lambda^N) \\
	\end{pmatrix}.
	\label{eq:P_N}
	\end{equation}
\end{widetext}
For the leading order error term, the differences between the three different $\lambda_{\parallel,\bot}^{(*)}$ and the dependency on the experimental parameters are neglected, and the product of any combination of eigenvalues is denoted by $\lambda^N$. Eq.~\eqref{eq:P_N} shows that all matrix elements except the first column approach zero for large $N$ since $|\lambda_{\parallel,\bot}^{(*)}| < 1$. This reveals that the population transfer between the individual transient eigenstates are negligible, and we are left with the population transfer from the steady eigenstate to the transient eigenstates, as described by the first column. Its entries describe the counteraction of populating the transient eigenstates, denoted by $P_{\text{S} \rightarrow f}(\kappa_n,\Delta \kappa_n)$ with $f \in \{\parallel, \bot, \bot^*\}$, and the relaxation of the transient eigenstates in the time span between their population and the time of observation after $N$ repetitions, denoted by $\prod_{k=n}^{N} \lambda_f(\kappa_k)$. 

The entries in the first column of Eq.~\eqref{eq:P_N} can be bound by 
\begin{equation}
\begin{split}
\left| \sum_{n=1}^{N} P_{\text{S} \rightarrow f}(\kappa_n,\Delta \kappa_n) \prod_{k=n}^{N} \lambda_f(\kappa_k) \right| \\
\leq 
\max_k \left| P_{\text{S} \rightarrow f}(\kappa_k,\Delta \kappa_k)  \sum_{n=0}^{N-1} \lambda_f^{n}(\kappa_k) \right| \\
\approx
\max_k \frac{\left| P_{\text{S} \rightarrow f}(\kappa_k,\Delta \kappa_k) \right|}{\left| 1 - \lambda_f(\kappa_k) \right|} .
\end{split}
\label{eq:bound_P_N}
\end{equation} 
Here, we used the geometric series
\begin{equation*}
\begin{split}
\sum_{n=0}^{N-1} \lambda_f^n = \frac{1 - \lambda_f^N}{1 - \lambda_f} \approx \frac{1}{1 - \lambda_f},
\end{split}
\end{equation*} 
where a large $N$ was assumed for the second step. 

In order to derive a limit under which we can neglect the individual transient eigenstates, we compare the corresponding elements of the first column in Eq. \eqref{eq:P_N} to the element corresponding to the steady-state eigenstate, which is one. Incorporating Eq.~\eqref{eq:bound_P_N}, this corresponds to the condition
\begin{equation}
\max_k \frac{\left| P_{\text{S} \rightarrow f}(\kappa_k,\Delta \kappa_k) \right|}{\left| 1 - \lambda_f(\kappa_k) \right|} \ll 1,
\label{eq:adiabaticity_def}
\end{equation} 
which ensures that the corresponding eigenstate is not populated. 

\vspace{0.15cm}
\subsubsection{The Hybrid State Adiabaticity Condition}
In this section, we will use the Taylor expansion in Eq.~\eqref{eq:P_elements} to solve Eq.~\eqref{eq:adiabaticity_def} for the cases of the perpendicular eigenstates, i.e. for $f = \bot^{(*)}$. 
Note that $P_{\text{S} \rightarrow \bot}$ and $P_{\text{S} \rightarrow \bot}^{*}$, as defined by Eq.~\eqref{eq:P_elements}, are complex conjugate of each other. 

Assuming that the eigenvectors are normalized to have a unit $\ell_2$-norm, we can bound the numerator of Eq.~\eqref{eq:P_elements} by
\begin{equation}
|\mathbf{u}_f^H(\kappa_n) \mathbf{U}'(\kappa_n) \mathbf{v}_d(\kappa_n)| \leq ||\mathbf{U}'||_2 \leq 1 .
\label{eq:uAv_bound}
\end{equation}
The here employed subordinate matrix norm is given by the square root of the largest eigenvalue of $(\mathbf{U}')^H \mathbf{U}'$ and is smaller than one since the $\mathbf{U}'$ consists only of rotations and relaxation terms (cf. Eq.~(53.5), Chapter~1 and Eq.~(8.4), Chapter~2 of Ref. \cite{Wilkinson1965}). 

The first term of the denominator in Eq.~\eqref{eq:P_elements}, $1 - \lambda_\bot^{(*)}$, describes the gap of the eigenvalues. We can assume $|\lambda_\bot^{(*)}| = 1$ as a worst case scenario and bound this gap by the complex phase $\Omega$. This gap can only be small when $\Omega$ approaches zero (Eqs.~\eqref{eq:lambda_orth}-\eqref{eq:lambda_orth_phase}), so that we can use a Taylor expansion of Eq.~\eqref{eq:lambda_orth_phase}
\begin{equation}
\text{Im} \{\tilde{\lambda}_\bot^{(*)} \}^2 \approx \sin^2 \frac{\alpha}{2} + \sin^2 \frac{\phi}{2}
\label{eq:Taylor_lamba}
\end{equation}
to derive the limit
\begin{equation}
|1 - \lambda_\bot^{(*)}| \geq \sqrt{\sin^2 \frac{\alpha}{2} + \sin^2 \frac{\phi}{2}}.
\label{eq:lambda_orth_bound}
\end{equation}

The last term in Eq.~\eqref{eq:P_elements} that requires our attention is $\mathbf{u}_\bot^H \mathbf{v}_\bot$. In order to assess the scenarios under which this product is small, we can approximate the evolution matrix by $\mathbf{U} = \mathbf{R} + \epsilon \mathbf{D} + \mathcal{O}(\epsilon^2)$, which views it as a small perturbation of the unitary rotation matrix $\mathbf{R} = \mathbf{R}_z \mathbf{R}_y \mathbf{R}_z$. The perturbation is of the order $\epsilon = 1 - \sqrt{E_2}$, and $\mathbf{D} = \{\mathbf{R}, \mathbf{C}\}$ is the anti-commuter of the rotation matrix and
\begin{equation}
\mathbf{C} = \begin{pmatrix}
-1 &                0 &               0 & 0 \\
0 &                -1 &               0 & 0 \\
0 &                0 &               -1 & 1 \\
0 &                0 &               0 &  0
\end{pmatrix},
\end{equation}
which approximates the relaxation matrix by $\mathbf{E} \approx \mathbb{1} + \epsilon \mathbf{C}$ when assuming $\delta \ll 1$. In this perturbation picture, the product of left- and right-eigenvectors $\mathbf{u}_f^H \mathbf{v}_f$ of the evolution matrix is approximated by
\begin{equation}
\mathbf{u}_f^H \mathbf{v}_f \approx 1 + \epsilon^2 \sum_{d \neq f} 
\frac{(\tilde{\mathbf{v}}_d^H \mathbf{D} \tilde{\mathbf{v}}_f)(\tilde{\mathbf{v}}_f^H \mathbf{D} \tilde{\mathbf{v}}_d)}{(\tilde{\lambda}_f - \tilde{\lambda}_d)^2},
\end{equation}
where the tilde indicates the eigenvalues and vectors of $\mathbf{R}$ (cf. Eq. \eqref{eq:P_elements} or Eq.~(10.2) in Chapter~2 of Ref. \cite{Wilkinson1965}). 
The first term results from the property $\tilde{\mathbf{u}}_f^H \tilde{\mathbf{v}}_f = 1$ of the eigenvectors of $\mathbf{R}$. Due to the orthornormality of the eigenspace of $\mathbf{R}$, we further eliminated the terms that are linear in $\epsilon$. With the bound $||\mathbf{D}||_2 \leq 1$ and the normalization of the eigenvectors, we follow $|\tilde{\mathbf{v}}_d^H \mathbf{D} \tilde{\mathbf{v}}_f| \leq 1$. Further, we can derive the eigenvalues of $\mathbf{R}$ from Eqs. \eqref{eq:lambda_par} and \eqref{eq:lambda_orth} by setting $E_1 = E_2 = 1$ and find $\tilde{\lambda}_S = \tilde{\lambda}_\parallel = 1$ and $\tilde{\lambda}_\bot^{(*)} = e^{\pm i \Omega}$. 
We adopt the bound in Eq.~\eqref{eq:lambda_orth_bound} for $d \in \{\text{S}, \parallel\}$ and for $d = \bot^*$ we find $|\tilde{\lambda}_\bot - \tilde{\lambda}_\bot^*|^2 \geq 2 (\sin^2 \frac{\alpha}{2} + \sin^2 \frac{\phi}{2})$
\footnote[4]{This bound neglects the scenario in which $\lambda_\bot^{(*)}$ both approach negative one, which is the case when $|\cos \frac{\alpha}{2}| \ll 1$ or $|\cos \frac{\phi}{2}| \ll 1$. Note that this leads to a breakdown of the approximations made for deriving Eq. \eqref{eq:lambda_orth}. Since both eigenvalues have the same complex phase, we can tread those two components jointly and without proof we state that both scenarios result in $||P_{\text{S} \rightarrow \bot}^{(1)} \mathbf{v}_\bot^{(1)} + P_{\text{S} \rightarrow \bot}^{(2)} \mathbf{v}_\bot^{(2)} ||_2 \ll 1$ where the superscript indicates the two formally complex conjugate components. In other words, when the eigenvalues $\lambda_\bot^{(*)}$ approach negative one, the perpendicular eigenstates are not populated.}. 
By summing over all three terms, we arrive at
\begin{equation}
	|\mathbf{u}_\bot^H \mathbf{v}_\bot | \geq 1 - \frac{5}{2} \frac{\epsilon^2}{\sin^2 \frac{\alpha}{2} + \sin^2 \frac{\phi}{2}}.
	\label{eq:uv_bound_Taylor}
\end{equation}

Inserting the bounds of the individual terms of the perturbation matrix (Eqs.~\eqref{eq:uAv_bound}, \eqref{eq:lambda_orth_bound}, and \eqref{eq:uv_bound_Taylor}) into Eq.~\eqref{eq:P_elements}, and using $1-E_2 \geq \epsilon^2$, we find 
\begin{equation}
	\left|P_{S \rightarrow \bot}^{(*)} (\alpha_n, \phi_n, \Delta \kappa_n) \right| \leq \Delta \kappa_n \frac{\sqrt{\sin^2 \frac{\alpha}{2} + \sin^2 \frac{\phi}{2}}}{\sin^2 \frac{\alpha_n}{2} + \sin^2 \frac{\phi_n}{2} - 5/2 (1-E_2)}.
	\label{eq:P_bound}
\end{equation}
This bound describes how much magnetization is at most transfered from the steady state to the orthogonal eigenstates by varying $\alpha$ or $\phi$ between two consecutive repetitions. 

Further, inserting into Eq.~\eqref{eq:adiabaticity_def} in order to account for the cumulative population, and utilizing Eq.~\eqref{eq:lambda_orth_bound}, we arrive at the limit 
\begin{equation}
\max_n |\Delta \kappa_n| \ll \sin^2 \frac{\alpha_n}{2} + \sin^2 \frac{\phi_n}{2} - \frac{5}{2} (1 - E_2).
\tag{\ref{eq:adiabaticity_HSFP}'}
\label{eq:adiabaticity_HSFP'}
\end{equation} 
When this adiabaticity condition is fulfilled, we can neglect the perpendicular transient eigenstates. 

\vspace{0.15cm}
\subsubsection{The Steady State Adiabaticity Condition}
In order to do the same analysis for the parallel transient eigenstate, we have to rely on the absolute value of $\lambda_\parallel$, since it is real-valued and positive. Note that $\mathbf{u}_\parallel^H \mathbf{v}_\parallel$ cannot be bound in the same way as done in Eq.~\eqref{eq:uv_bound_Taylor} since the eigenvalues $\tilde{\lambda}_S = \tilde{\lambda}_\parallel$ are degenerate. Since the adiabaticity condition of the parallel eigenstate is not essential for this work, we skip the degenerate perturbation theory and assume $\mathbf{u}_\parallel^H \mathbf{v}_\parallel \approx 1$. With the bound $\lambda_\parallel \leq E_1$, which result from Eq.~\eqref{eq:lambda_par}, and with Eqs.~\eqref{eq:adiabaticity_def}-\eqref{eq:uAv_bound}, we arrive at the adiabaticity condition
\begin{equation}
|\Delta \kappa_n | \ll (1 - E_1)^2,
\tag{\ref{eq:adiabaticity_SSFP}'}
\end{equation} 
which ensures that the parallel transient state is negligible. 

\subsection{The Bloch Equation in Spherical Coordinates}
\label{sec:Bloch_Spherical}
Under the derived adiabaticity condition, the hybrid state emerges, and we observe transient-state behavior only along the direction of the steady-state magnetization. Transforming the Bloch equation into spherical coordinates isolates the transient-state behavior in a single dimension, and the components of the Bloch equation uncouple into first order differential equations that can be solved. 

Spherical coordinates are here defined by $x = r \sin{\vartheta} \cos \varphi$, $y = r \sin{\vartheta} \sin \varphi$ and $z = r \cos{\vartheta}$, where $r$ is the radius, $\vartheta$ the polar angle or the angle between the magnetization and the $z$-axis and $\varphi$ is the azimuth or the angle between the $x$-axis and the projection of the magnetization onto the $x$-$y$-plane. In order to better highlight effect of inversion pulses, we use the limits $-1 \leq r \leq 1$, $0 \leq \vartheta \leq \pi/2$, and $0 \leq \varphi < 2 \pi$ to uniquely identify the polar coordinates.
Thermal equilibrium is given by $r_0=1$, $\vartheta_0=0$ and $\varphi_0=0$, where the latter can be chosen freely. 

Since the azimuth, or phase, adiabatically transitions between steady states, we can transform the known Cartesian steady-state solutions (Eqs. (6,7) in Ref.~\cite{Freeman1971}) to spherical coordinates, which results in Eq.~\eqref{eq:varphi}. The polar angle can be derived from Eqs. (9-11) in Ref.~\cite{Freeman1971} and is given by
\begin{equation}
	\tan \vartheta = \frac{\sqrt{E_2} \sin \alpha \sqrt{1 - 2 E_2 \cos \phi + E_2^2}}{G + \sqrt{E_1} (E_2 (E_2-\cos \phi) + (1 - E_2 \cos \phi) \cos \alpha)}
	\label{eq:vartheta_exact}
\end{equation}
with
\begin{equation*}
\begin{split}
	G = &\frac{(1 - E_1 \cos \alpha) (1 - E_2 \cos \phi)}{1 + \sqrt{E_1}} \\
	    &-\frac{(E_1 - \cos \alpha) (E_2 - \cos \phi) E_2}{1 + \sqrt{E_1}}.
\end{split}
\end{equation*}
With a Taylor expansion at $E_2 = 1$, the polar angle is described by
\begin{equation}
\begin{split}
\sin^2 \vartheta = &\frac{\sin^2 \frac{\alpha}{2}}{\sin^2 \frac{\phi}{2}  \cdot \cos^2 \frac{\alpha}{2} + \sin^2 \frac{\alpha}{2}} \\ &+ (1-E_2) \cdot \xi + \mathcal{O} ((1-E_2)^2)
\end{split}
\label{eq:vartheta_c}
\end{equation}
with
\begin{equation*}
\xi = \frac{4 (\cos \alpha - 1)^2 (\sqrt{E_1} - 1)}{(\sqrt{E_1} + 1) (\cos\alpha + \cos\phi + \cos\alpha \cos\phi - 3)^2}.
\end{equation*}
The factor $\xi$ is only large, if $\cos\phi \approx (3-\cos\alpha)/(\cos\alpha+1)$, which is only the case, if $|1-\cos\alpha| \ll 1$ and $|1 - \cos\phi| \ll 1$ are simultaneously fulfilled, i.e. for small flip angle and in the vicinity of the stop-band. Consequently, for standard imaging scenarios with $T_R \ll T_2$ the polar can be approximated by Eq.~\eqref{eq:vartheta} apart from the vicinity of the stop band. 

The spherical coordinate $r$ captures the transient-state spin dynamics, and we can derive Eq.~\eqref{eq:Bloch_2D} simply by transforming the Bloch equation into spherical coordinates\cite{Tahayori2009,Lapert2013}.

\subsection{$B_1$-inhomogeneities}
One can describe the effect of $B_1$-inhomogeneities on the spins by $\alpha = B_1/B_1^{\text{nom.}} \alpha^{\text{nom.}}$, where $B_1^{\text{nom.}}$ and $\alpha^{\text{nom.}}$ describe the nominal $B_1$-field and flip angle, respectively. The effect on the polar angle is described by inserting this relation into Eq. \eqref{eq:vartheta} and successively into Eq. \eqref{eq:r}. 

In order to implement anti-periodic boundary conditions, the magnetization must be inverted between successive cycles ($r(0) = - r(T_C)$), while changes of $\vartheta$ and $\varphi$ are required to remain within limits in order not to violate the adiabaticity condition posed in Eq.~\eqref{eq:adiabaticity_HSFP}. Applying a $\pi$-pulse with an inhomogeneous $B_1$-field would lead to severe fluctuations of $\vartheta$, causing a violation of the adiabaticity condition. In order to mitigate these fluctuations, we surround the inversion pulse by crusher gradients. As shown in Refs. \cite{Hennig1991,Weigel2015}, the transversal magnetization $M_\perp$ refocuses after inversion pulse with crusher gradients to an echo of the size $M_\perp^+ = \sin^2 (\pi/2 \cdot B_1/B_1^{\text{nom.}}) M_\perp^-$, where the superscript $+$ and $-$ indicate the magnetization before and after the RF pulse, respectively. The longitudinal magnetization, on the other hand, is given by $M_z^+ = \cos (\pi B_1/B_1^{\text{nom.}}) M_z^-$. In spherical coordinates, this leads to 
\begin{equation}
\tan \vartheta^+ = \frac{\sin^2 (\frac{\pi}{2} \frac{B_1}{B_1^{\text{nom.}}})}{\cos (\pi \frac{B_1}{B_1^{\text{nom.}}})} \tan \vartheta^-.
\label{eq:delta_vartheta}
\end{equation}
In the human brain at 3T, one usually observes variations in the range of $B_1/B_1^{\text{nom.}} \in [0.8, 1.2]$ \cite{Chung2010}. Within this range, the resulting effect is bound by $|\vartheta^+/\vartheta^- - 1| < 0.12$ and will be neglected in the following. 

In return, the crusher gradients manipulate $r$, which is accounted for by 
setting
\begin{equation}
\beta = - \sqrt{ \sin^2 \vartheta^- \cdot \sin^4 \frac{\pi B_1}{2 B_1^{\text{nom.}}} + \cos^2 \vartheta^- \cdot \cos^2 \frac{\pi B_1}{B_1^{\text{nom.}}}}
\label{eq:beta_1}
\end{equation}
in Eq.~\eqref{eq:r}.
Repeating the inversion pulses with the same spoiling gradients can potentially result in higher order spin echoes and stimulated echoes, impairing the derived description of the spin physics. However, when using $T_C \gg T_2$, we can assume that those contributions are negligible. 

\subsection{Numerical Optimizations}
\subsubsection{Cram\'er Rao Bound}
The Cram\'er-Rao bound\cite{Rao1945,Cramer1946} provides a universal limit for the noise variance of a measured parameter, given that the reconstruction algorithm is an unbiased estimator. This very general and established metric has been utilized for optimizing MR parameter mapping experiments in Refs. \cite{Jones1996,Jones1997,Teixeira2017} amongst others, and to MRF in particular in Ref. \cite{Zhao2016b}. In discretized notation, the Cram\'er-Rao bound is defined by the inverse of the Fisher information matrix $\mathbf{F}$ with the entries $\mathbf{F}_{ij}=\mathbf{b}_i^T\mathbf{b}_j/\sigma^2$ given by 
\begin{align*}
\mathbf{b}_1 &= d\mathbf{x}/dPD\\ 
\mathbf{b}_2 &= d\mathbf{x}/dT_1 \\ 
\mathbf{b}_3 &= d\mathbf{x}/dT_2.
\end{align*}
Here $\mathbf{x} \in \mathbb{R}^{N_t}$ is a vector describing the measured signal or, equivalently, the transversal magnetization at $N_t$ discrete time points, and $\sigma^2$ is the input variance. Each element of the vector is given by $x_n = r(t_n) \cdot \sin \vartheta(t_n)$. The vectors $\mathbf{b}_i$ describe the derivatives of the signal evolution with respect to all considered parameters. Note that the proton density is here normalized to $PD=1$, so that $\mathbf{b}_1 = \mathbf{x}$. 

In this work, we focused on quantifying relaxation times, since $PD$, as defined in this work, is modulated by the receive coil sensitivity and provides only a relative measure. We can define the dimensionless relative Cram\'er-Rao bounds to be
\begin{align}
rCRB(T_1) &= \frac{1}{\sigma^2 T_1^2} \frac{T_C}{T_R} (\mathbf{F}^{-1})_{2,2}
\label{eq:rCRB_T1} \\
rCRB(T_2) &= \frac{1}{\sigma^2 T_2^2} \frac{T_C}{T_R} (\mathbf{F}^{-1})_{3,3}.
\label{eq:rCRB_T2}
\end{align}
The normalization by the variances cancels out the variance in the definition of the Fisher information matrix, and the normalization by the relaxation time is done to best reflect the $T_{1,2}$-to-noise ratio (defined as $T_{1,2}/\sigma_{T_{1,2}}$). Further, the multiplication with $T_C/T_R$ normalizes the $rCRB$ by duration of the experiment such that it can be understood as the squared inverse SNR efficiency per unit time, given a fixed $T_R$. 

\vspace{0.15cm}
\subsubsection{Optimal Control}
The polar angle $\vartheta$ is here treated as the control parameter for spin dynamics along the radial direction as by Eq.~\eqref{eq:Bloch_2D}. Thus, we can employ the rich optimal control literature \cite{Conolly1986,Skinner2003} for numerical optimization of $\vartheta(t)$. We used a Broyden-Fletcher-Goldfarb-Shanno (BFGS) algorithm \cite{DeFouquieres2011} with $rCRB(T_1) + rCRB(T_2)$ as an objective function. To further improve convergence, the BFGS algorithm is embedded in a scatter search algorithm which tried 1000 starting points \cite{Ugray2007}. The numerical optimization was based on $\vartheta(\Delta t\cdot n)$ with a discrete step size of $\Delta t = 4.5~\text{ms}$ and the evaluation points $n \in \{1, 2, \ldots, T_C/T_R\}$. The gradient of the objective function with respect to $T_1$, $T_2$, and each $\vartheta(\Delta t\cdot n)$ was explicitly calculated. 

Since the $rCRB$ intrinsically compares a signal evolution to its surrounding in the parameter space, only a single set of relaxation times is necessary for the optimization. Here, we used the relaxation times $T_1={781}~\text{ms}$ and $T_2={65}~\text{ms}$, corresponding to the values measured for white matter as reported in Ref. \cite{Jiang2015}. All optimizations were initialized with the pattern provided in the pSSFP paper \cite{Asslander2017} and the optimizations were performed with the constraint $0\leq\vartheta\leq\pi/4$, which limits the flip angle to $\alpha\leq\pi/2$, ensuring consistent slice profiles by virtue of the linearity in the small tip-angle approximation \cite{Hoult1979}, and aiding compliance with safety considerations by avoiding high power large flip-angle pulses.

\subsection{In Vivo Experiments}
An asymptomatic volunteer's brain was imaged following written informed consent and according to a protocol approved by our institutional review board. A measurement was performed with the anti-periodic bHSFP experiment
on a 3T Prisma scanner (Siemens, Erlangen, Germany). The 16 head elements of the manufacturer's 20 channel head/neck coil were used for signal reception.

Spatial encoding was performed with a sagittally oriented 3D stack-of-stars trajectory, which starts at the outer k-space and acquires for one $T_C$ data while incrementing the angle of the k-space spoke by twice the golden angle increment \cite{Winkelmann2007}. These large gaps are filled by repeating this procedure one time with the entire k-space trajectory rotated by the golden angle. Thereafter, the next 3D phase encoding step is performed in the exact same way, while adhering to the Nyquist-Shannon theorem along the slice direction. The acquired resolution of the maps is $1~\text{mm} \times 1~\text{mm} \times 2~\text{mm}$ at a FOV of $256~\text{mm} \times 256~\text{mm} \times 192~\text{mm}$. The readout dwell time was set to $2.1~\upmu\text{s}$ and an oversampling factor of 2 was applied. We used a $T_R = 4.5$ms and the readout was skipped in segments with a polar angle close to zero (gray areas in supporting Fig.~\ref{fig:Spin_Dynamics}), so that 601 spokes were acquired during one $T_C$. The total scan time was approximately $12.24~\text{min}$. 

Along the fully sampled phase encoding direction, a Fourier transformation was performed and, thereafter, each slice was treated separately. The raw data were compressed to 8 virtual receive coils via SVD compression \cite{Huang2008}, followed by image reconstruction with the low rank alternating direction method of multipliers (ADMM) approach proposed in Ref.\cite{Asslander2018}, which includes parallel imaging \cite{Sodickson1997,Pruessmann2001,Uecker2014}. The data consistency step of the ADMM algorithm was performed with 20 conjugate gradient steps. In order to prevent non-linear effects from impairing the noise assessment, only a single ADMM iteration was performed and no spatial regularization was applied.

The employed dictionaries include the parameter values $T_1 (\text{s}) = 0.1 \cdot 1.01^j \; \forall \; j \in \{0, 1, \ldots, 413\}$, thus covering the range between $100$~ms and $6$~s in steps of $1$\%. The dictionaries covered the range of $T_2$ values between $10$~ms and $3$~s in steps of $1$\%, i.e. $T_2 (\text{s}) = 0.01 \cdot 1.01^j \; \forall \; j \in \{0, 1, \ldots, 575\}$. The dictionaries further discretized $\phi \in [0, \pi]$ into 15 bins and $B_1/B_1^{\text{nom.}} \in [0.8, 1.2]$ into 40 bins. The dictionary was compressed to include the singular vectors corresponding to the 12 largest singular values resulting from a singular value decomposition of the dictionary matrix \cite{McGivney2014}. 

In the matching step of each voxel, only fingerprints were considered that matched the $\phi$ and $B_1/B_1^{\text{nom.}}$ from separate scans. The $\phi$ map was acquired with a double-echo SPGR experiment and the $B_1$ map with a turboFLASH experiment, as described in Ref.\cite{Chung2010}. 

\section{Code availability}
The source code used for the current study is available from the corresponding author on reasonable request.

\section{Data availability}
The datasets generated and analyzed during the current study are available from the corresponding author on reasonable request.

\section{Author Contributions}
JA and DSN derived the theory. JA performed the numerical optimizations, simulations and the experiment. JA, RL and MAC analyzed and interpreted the data. DKS provided consultancy. JA wrote the paper with the help of all authors. All authors have critically reviewed the manuscript.

\section{Acknowledgements}
The authors would like to thank Steffen Glaser, Quentin Ansel and Dominique Sugny for fruitful discussions, and for giving insights into their optimal control implementation. The authors would also like to acknowledge Jeffrey Fessler and Gopal Nataraj for discussions regarding the solution of the simplified Bloch equation.

This work was supported by the research grants NIH/NIBIB R21 EB020096 and NIH/NIAMS R01 AR070297, and was performed under the rubric of the Center for Advanced Imaging Innovation and Research (CAI2R, www.cai2r.net), a NIBIB Biomedical Technology Resource Center (NIH P41 EB017183).

\bibliographystyle{naturemag}
\bibliography{library}

\end{document}